\documentclass[fleqn,10pt]{wlscirep}

\usepackage[utf8]{inputenc}
\usepackage[T1]{fontenc}
\usepackage{multirow}
\usepackage{placeins}
\usepackage{textcomp}
\usepackage{float}
\usepackage{subcaption}
\usepackage{caption}
\usepackage{amsmath}
\usepackage{booktabs}
\usepackage{xurl}
\usepackage{listings}
\usepackage{xcolor}
\usepackage[many]{tcolorbox}
\tcbuselibrary{listings,breakable,skins}
\lstdefinestyle{PromptBox}{
  basicstyle=\footnotesize\ttfamily,
  breaklines=true,
  breakatwhitespace=false,
  columns=fullflexible,
  keepspaces=true,
  showstringspaces=false,
  aboveskip=0pt,
  belowskip=0pt,
  frame=none,
  backgroundcolor=\color{gray!5}
}
\newtcolorbox{PromptFrame}[1]{%
  enhanced,
  breakable,
  sharp corners,
  boxrule=0pt,
  colframe=white,
  colback=gray!5,
  colbacktitle=black!50,
  coltitle=black,
  fonttitle=\bfseries\sffamily\footnotesize,
  title={#1},
  left=6pt,
  right=6pt,
  top=4pt,
  bottom=4pt,
  toptitle=3pt,
  bottomtitle=3pt,
  before skip=0.7em,
  after skip=0.85em
}

\newcommand{\TextMol}{Text$\to$Mol}
\newcommand{\MolText}{Mol$\to$Text}

\newcommand{\methodheading}[1]{\par\vspace{0.35em}\noindent\textbf{#1}\par\vspace{0.15em}}

\newcounter{sinote}
\renewcommand{\thesinote}{S\arabic{sinote}}
\newcommand{\sinote}[2]{%
  \refstepcounter{sinote}\label{#1}%
  \subsection*{Note~\thesinote. #2}%
}

\title{A local orchestration harness for molecular design with foundation models}

\author[1, **]{Sakib Mostafa}
\author[1, **]{Md Atik Ahamed}
\author[2,3]{James Zou}
\author[1,4,5]{Lei Xing}
\author[1, *]{Md Tauhidul Islam}

\affil[1]{Department of Radiation Oncology, Stanford University, Stanford, California, USA}
\affil[2]{Department of Computer Science, Stanford University, Stanford, California, USA}
\affil[3]{Department of Biomedical Data Science, Stanford University, Stanford, California, USA}
\affil[4]{Institute for Computational and Mathematical Engineering, Stanford University, Stanford, California, USA}
\affil[5]{Department of Electrical Engineering, Stanford University, Stanford, California, USA}

\affil[**]{These authors contributed equally: Sakib Mostafa and Md Atik Ahamed}
\affil[*]{Correspondence: tauhid@stanford.edu}

\begin{abstract}
Molecular design requires coordinated generation, property prediction and iterative editing, with each decision linked to the correct chemical structure and supporting evidence. Integrating these capabilities in practice presents three major challenges: retaining local control of molecular data, connecting specialized models across tasks and verifying that proposed changes satisfy chemical constraints. Here we present MolCrew, a locally deployable system that coordinates four separately trained molecular specialists with a language-model orchestrator and RDKit calculations. The specialists generate structures from descriptions, describe molecular structures, predict biological endpoints and propose edits towards specified goals. Separate model weights allow each capability to be evaluated and updated independently within the workflow. RDKit calculations check structural validity, quantify descriptor changes and assess specified structural constraints, while returned observations guide further tool calls. We evaluated low-rank adaptation of the orchestrator for tool-use decisions and failure recovery while leaving specialist weights unchanged. On 3,300 ChEBI-20 test examples, MolCrew achieved 96.4\% exact structure recovery, compared with 38.7\% for published Atomas-large results. Across eight MoleculeNet datasets evaluated using scaffold splits, mean area under the receiver operating characteristic curve was 0.851, compared with 0.770 for published Atomas results. In a recorded evaluation using the unadapted orchestrator, 1,554 of 1,699 requests (91.5\%) passed every predefined workflow check. By preserving inspectable links between molecular proposals, model outputs and chemical calculations, MolCrew provides a modular framework for computational candidate assessment and refinement under local data control.
\end{abstract}

\begin{document}

\flushbottom
\maketitle
\thispagestyle{empty}

\section{Introduction}\label{sec:intro}
Molecular discovery requires repeated decisions about which compounds to synthesize, how to assess them and what to change next. Activity at a biological target must be balanced against solubility, permeability, metabolic stability and potential toxicity, which together influence a compound’s prospects for development~\cite{hughes2011principles,paul2010improving,lipinski2001experimental}. Because a structural change can affect several properties, improving one feature requires reassessing the molecule as a whole. Computational methods support this iterative process through molecular generation, property prediction and optimization~\cite{vamathevan2019applications}. Integrating these capabilities into practical workflows presents three challenges: retaining local control of molecular data, coordinating specialized models across tasks and verifying that proposed changes satisfy specified chemical constraints. As candidates are generated, evaluated and refined, each structure must remain linked to its supporting evidence and the objective guiding its modification.

Local data control is particularly relevant when workflows involve proprietary molecular structures and experimental measurements, whose confidentiality has motivated privacy-preserving approaches in pharmaceutical research~\cite{heyndrickx2024melloddy}. Local language-model execution has also been demonstrated in chemical applications, including the interpretation of structure–property relationships~\cite{wellawatte2025human} and the coordination of robotic experiments~\cite{song2025multiagent}. For workflows in which structures, design objectives and intermediate results must remain within a locally controlled computing environment, the deployment requirement extends to every component that processes these data. This includes the specialist models, the coordinating model and the tools used to evaluate proposed structures, making deployment and coordination closely connected design considerations.

Foundation models provide reusable components for these workflows by transferring information learned during pretraining across molecular prediction and generation tasks~\cite{raffel2020exploring,edwards2022molt5}. MolT5 connects molecular structures with natural-language descriptions, enabling translation between the two representations~\cite{edwards2022molt5}. Multitask language modeling extends this approach to a broader range of chemical and textual tasks~\cite{christofidellis2023unifying}, while other methods align molecular and textual representations for retrieval, editing, understanding and generation~\cite{liu2023moleculestm,zhang2025atomas}. Together, these advances allow researchers to express molecular objectives in language and act on them computationally. Integrating these capabilities into iterative design requires maintaining a consistent molecular context as structures and observations pass between models.

This coordination becomes particularly important when a molecule changes during a workflow. Subsequent predictions and calculations must refer to the edited structure, and each output must be interpreted according to what it measures. Exact structure recovery, for example, concerns molecular identity, whereas a predicted biological endpoint depends on the assay, training data and evaluation setting~\cite{edwards2022molt5,wu2018moleculenet,yang2019analyzing}. A request to increase a hydrogen-bond count while retaining a molecular core therefore requires both numerical and structural checks. RDKit provides structure parsing, molecular comparisons and descriptor calculations that can supply this feedback~\cite{landrum2013rdkit}, and methods such as DrugAssist use feedback to guide successive edits~\cite{ye2025drugassist}. These checks establish whether specified chemical constraints have been met; an improvement in a calculated descriptor does not establish improved biological activity~\cite{bemis1996scaffold,bickerton2012qed}. A reliable workflow must track the current structure, its associated observations and the conditions that remain unsatisfied.

Language-model agents offer a way to organize these steps through a common interface. ChemCrow combines language models with chemistry tools, while Co-scientist connects planning, information retrieval, computation and laboratory experiments~\cite{bran2024chemcrow,boiko2023coscientist}. Broader scientific agent systems coordinate specialized components across stages of investigation~\cite{ghareeb2026robin}. For molecular design, evaluating such systems requires separating the capabilities of individual models from the decisions that connect them. A failed request may arise from an unsuitable molecular proposal, an inappropriate tool choice or a calculation associated with the wrong structure. Assessing specialists and complete workflows separately helps identify these sources of error and establish where coordination contributes to task completion.
The coordinating model can itself be trained to select tools, supply appropriate arguments and respond to unsuccessful calls. Low-rank adaptation enables this training while leaving the pretrained model parameters unchanged~\cite{hu2022lora}, allowing coordination to be adapted independently of the molecular specialists. Its evaluation must address both decision quality and task completion. Tests with supplied observations assess whether the orchestrator chooses an appropriate next action. Workflows that execute the tools assess whether those decisions remain useful when they determine the observations received next. This distinction matters because a valid tool call can still produce a molecule that misses the requested target, and a useful molecular proposal can still be accompanied by an incomplete response.

\begin{figure}[pt]
\centering
\includegraphics[width=\linewidth]{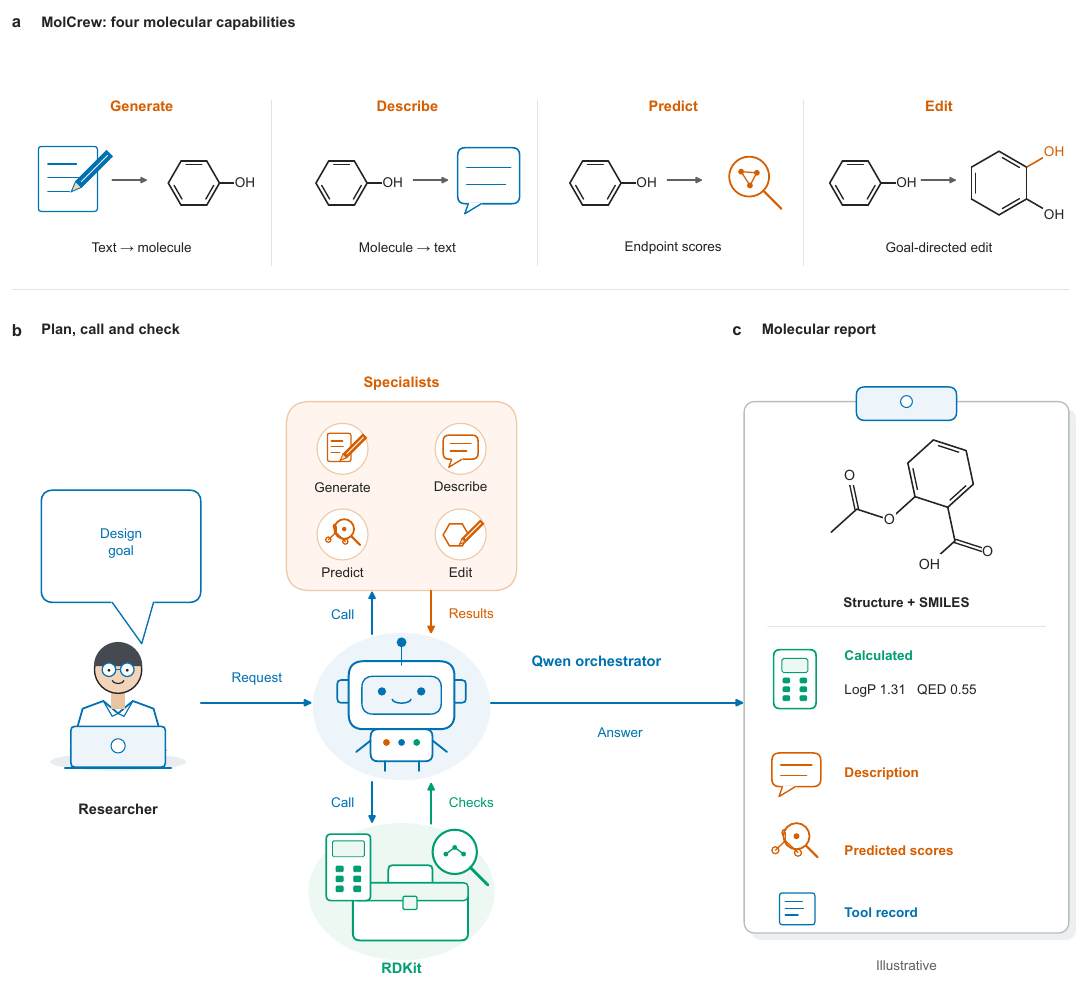}
\caption{\textbf{MolCrew combines molecular specialists with tool-guided planning.} \textbf{a},~Four separately trained specialists support complementary molecular tasks: generating candidate structures from descriptions, describing structures in language, predicting molecular endpoint scores, and proposing edits to a seed structure under a stated goal. The specialists use a shared backbone architecture and separate model weights. \textbf{b},~A pretrained Qwen orchestrator uses the user request, molecular context and prompt instructions to select specialist or RDKit calls. Specialist results and chemistry-check observations inform subsequent planning within a bounded call budget. Descriptor requests can use RDKit directly, and an individual request need not invoke all four specialists. The original orchestrator uses prompt refinement without weight training. RDKit supports structure parsing, depiction, descriptor calculation and checks of specified descriptor targets. \textbf{c},~A response can include the requested molecular structure and SMILES, generated descriptions, learned endpoint scores and calculated descriptors, accompanied by recorded tool calls and check results. Aspirin illustrates the structure and calculated-descriptor fields (RDKit LogP, 1.31; QED, 0.55). Molecular examples in both \textbf{a} and \textbf{c} are illustrative rather than observed model predictions or a recorded successful episode. Learned endpoint scores are distinct from calculated descriptors. Chemistry checks do not establish biological activity, caption factuality or satisfaction of every natural-language constraint. QED, quantitative estimate of drug-likeness; SMILES, simplified molecular-input line-entry system.}
\label{fig:overview}
\end{figure}

Here we present MolCrew, a locally deployable system that connects four molecular specialists with a language-model orchestrator and RDKit calculations (Fig.~\ref{fig:overview}). The specialists generate structures from descriptions, describe molecules, predict molecular endpoints and edit structures towards specified goals. They share a backbone architecture but have separate weights and task-specific training. We trained separate low-rank Qwen adapters for tool-use decisions and failure recovery while keeping the original Qwen parameters and specialist weights fixed, and retained the unadapted orchestrator as a reference. User requests and accumulated observations guide subsequent calls within a bounded budget. Local deployment supports control over molecular data, the orchestrator coordinates specialized capabilities across tasks, and RDKit calculations assess whether proposed structures satisfy specified chemical constraints. By recording molecular structures, calculations and model outputs, MolCrew makes these workflows inspectable and provides a framework for evaluating both molecular capabilities and the coordination that connects them.

\section{Results}\label{sec:results}
We evaluated MolCrew at three levels: individual molecular capabilities, tool-use decisions and completion of molecular workflows. Specialist benchmarks assessed structure recovery, agreement with reference captions, endpoint ranking and attainment of requested descriptor targets~\cite{edwards2022molt5,wu2018moleculenet,ye2025drugassist}. Using the same editing checkpoint, we examined whether inference-time feedback improved descriptor attainment. We assessed orchestration through the original serving evaluation with unadapted Qwen and separate matched studies of orchestrator adaptation and candidate handling. These studies examined whether improved agreement with prescribed decisions in synthetic contexts transferred to live molecular tasks, and whether suitable molecules generated by the editing specialist were successfully delivered in the final response. Molecular outcomes and completion of the required response were evaluated separately, using the success criteria specified for each analysis.

\begin{figure}[pt]
\centering
\includegraphics[width=\linewidth]{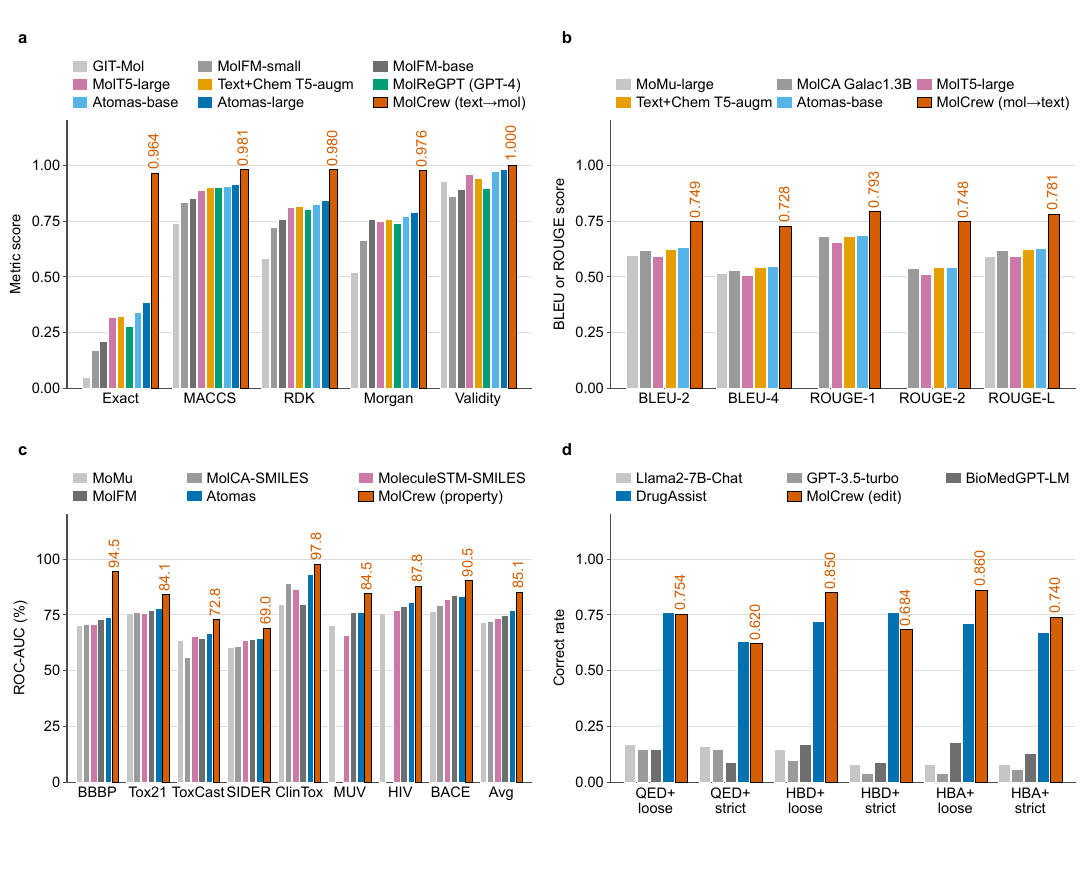}
\caption{\textbf{Performance across four molecular tasks.}
MolCrew (orange) is compared with published baseline results under each task's evaluation protocol. Complete numerical comparisons are provided in Supplementary Tables~\ref{tab:S1}--\ref{tab:S4}.
\textbf{a},~Structure generation from ChEBI-20 descriptions, assessed by exact structure recovery, molecular fingerprint similarity and RDKit validity~\cite{edwards2022molt5,zhang2025atomas}.
\textbf{b},~Molecular description generation on ChEBI-20, assessed by BLEU and ROUGE scores using SciBERT tokenisation~\cite{zhang2025atomas}.
\textbf{c},~Area under the receiver operating characteristic curve (AUROC) across eight MoleculeNet classification datasets evaluated using scaffold splits~\cite{wu2018moleculenet,zhang2025atomas}.
\textbf{d},~Success rates for instruction-directed edits that increase drug-likeness (QED), hydrogen-bond donor count (HBD) or hydrogen-bond acceptor count (HBA), evaluated under the loose and strict DrugAssist criteria~\cite{ye2025drugassist}. Editing training included pairs derived from benchmark test molecules, as detailed in Methods.}
\label{fig:main}
\end{figure}

\subsection{Molecular structures from language}\label{sec:textmol}

We evaluated how accurately \TextMol{} could recover molecular structures from 3{,}300 ChEBI-20 test descriptions, using published measures of exact recovery, fingerprint similarity and RDKit validity~\cite{edwards2022molt5,zhang2025atomas,hastings2016chebi,landrum2013rdkit}. Exact recovery, assessed by International Chemical Identifier (InChI) identity, reached 96.4\%, compared with the published Atomas-large result of 38.7\% (Fig.~\ref{fig:main}a; Supplementary Table~\ref{tab:S1}). RDKit validity was 100\% under the primary scoring protocol. Mean Morgan fingerprint similarity was 0.976, compared with 0.788 for Atomas-large. This measure captures structural agreement between predictions and references, including cases in which the structures are not identical. Expanding the caption--molecule training corpus from the ChEBI+PCdes base to a broader public-data mixture increased exact recovery from 84.0\% to 96.4\% (Supplementary Fig.~\ref{fig:si_textmol_ablation}).

We next compared calculated properties of the predicted and reference molecules. Among the 3{,}250 outputs that passed the parsing criteria for this analysis, mean quantitative estimate of drug-likeness (QED) was 0.379 for both groups. Mean Crippen LogP was 1.296 for predictions and 1.238 for references (Supplementary Note~\ref{note:textmol-fidelity}). The 50 outputs excluded from these summaries remained in the primary evaluation denominator. These results describe average calculated properties; they do not establish biological activity or experimental suitability.

\begin{figure}[pt]
\centering
\includegraphics[width=\linewidth]{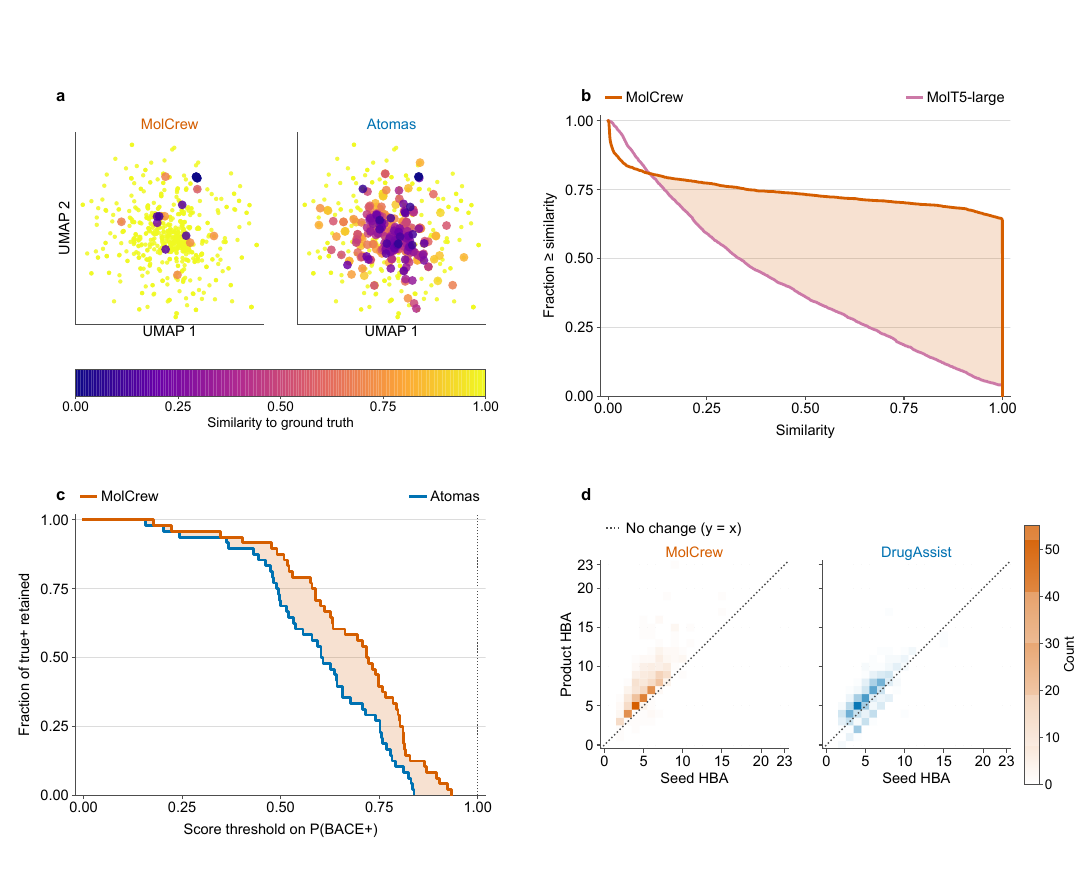}
\caption{\textbf{Comparisons of individual predictions across four molecular tasks.}
Local comparator analyses complement the published benchmarks in Fig.~\ref{fig:main}. Analysis methods and selection criteria are detailed in Supplementary Note~\ref{note:geometry-construction} and Supplementary Table~\ref{tab:S8}.
\textbf{a},~Uniform manifold approximation and projection (UMAP) of Morgan fingerprints for ChEBI-20 predictions, coloured by Tanimoto similarity to the reference: MolCrew (left) and local Atomas outputs (right). The plots show 650 examples from a shared embedding. Mean similarities of 0.978 and 0.776, respectively, use all 3{,}230 pairs with valid fingerprints; 70 pairs were excluded.
\textbf{b},~Fraction of captions at or above each term frequency--inverse document frequency (TF--IDF) cosine-similarity threshold relative to reference descriptions. Across 3{,}300 examples, mean similarity was 0.741 for MolCrew and 0.409 for MolT5-large. These diagnostic scores differ from the primary SciBERT-token scores.
\textbf{c},~Fraction of labelled positives retained as the prediction-score threshold increases on BACE, a dataset of human beta-secretase 1 inhibitors. The local Atomas comparator uses a logistic readout fitted to unchanged Atomas representations. The MolCrew and comparator sets contain 152 and 151 examples, respectively, including 48 positives each. The rationale for selecting BACE is given in Supplementary Note~\ref{note:geometry-construction}.
\textbf{d},~Starting and returned hydrogen-bond acceptor (HBA) counts under the loose increase criterion for MolCrew and local DrugAssist retrieval feedback. The diagonal indicates no change. Plots contain 455 and 499 structures, respectively; success rates use 500 requests per method.}
\label{fig:geometry}
\end{figure}

We also compared individual predictions with locally available Atomas outputs (Fig.~\ref{fig:geometry}a). Across 3{,}230 pairs with valid fingerprints, mean Morgan Tanimoto similarity to the reference was 0.978 for MolCrew and 0.776 for Atomas. This analysis excluded 70 pairs for which a required fingerprint was unavailable. Fingerprint identity in this subset is distinct from the primary InChI-based measure of exact recovery. Matched examples illustrate successful recoveries and a case missed by both methods; additional examples show differences in substituents, chain length and charge (Supplementary Figs~\ref{fig:si_case_textmol} and~\ref{fig:si_mismatch}). Selection procedures and analysis subsets are detailed in Supplementary Notes~\ref{note:case-selection} and~\ref{note:geometry-construction} and Supplementary Table~\ref{tab:S8}.

\subsection{Molecular descriptions from structures}\label{sec:moltext}

We evaluated the descriptions generated by \MolText{} against ChEBI-20 references using the SciBERT WordPiece scoring protocol adopted by Atomas~\cite{edwards2022molt5,zhang2025atomas,beltagy2019scibert}. BLEU-2 reached 74.9\%, compared with the published Atomas-base score of 63.2\%, and BLEU-4 reached 72.8\%, compared with 54.9\%. ROUGE-L was 78.1\%, compared with 62.6\%, with higher ROUGE-1 and ROUGE-2 scores also observed (Fig.~\ref{fig:main}b; Supplementary Table~\ref{tab:S2})~\cite{papineni2002bleu,lin2004rouge}. These scores indicate closer textual agreement with the reference descriptions.

For comparisons of individual molecules, we used available MolT5-large outputs because suitable Atomas captions were unavailable. Across 3{,}300 examples, mean term frequency--inverse document frequency (TF--IDF) cosine similarity to the reference was 0.741 for MolCrew and 0.409 for MolT5-large. MolCrew achieved higher similarity on 73.3\% of examples (Fig.~\ref{fig:geometry}b). Survival curves and matched descriptions illustrate how agreement varied across molecules (Supplementary Figs~\ref{fig:si_moltext_survival} and~\ref{fig:si_case_moltext}--\ref{fig:si_case_moltext_c}). TF--IDF similarity is a separate diagnostic from the primary SciBERT-token scores; neither measure alone establishes the factual accuracy of every generated statement (Supplementary Note~\ref{note:board-metrics}).

\subsection{Molecular endpoint prediction for candidate ranking}\label{sec:property}

We evaluated the property specialist on eight MoleculeNet classification datasets covering permeability, toxicity, adverse effects and biological activity, using the scaffold splits described in Methods~\cite{wu2018moleculenet,yang2019analyzing}. The unweighted mean area under the receiver operating characteristic curve (AUROC) was 85.1\%, compared with 77.0\% for published Atomas results, with higher scores on all eight datasets (Fig.~\ref{fig:main}c; Supplementary Table~\ref{tab:S3}). The largest differences were on the blood--brain barrier penetration (BBBP) dataset, where AUROC was 94.5\% for MolCrew and 73.7\% for Atomas, and the Maximum Unbiased Validation (MUV) screening benchmark, where the corresponding values were 84.5\% and 76.3\%. Comparisons across all endpoints are shown in Supplementary Fig.~\ref{fig:si_auroc_delta}.

Both methods had their lowest AUROC values on SIDER and ToxCast. MolCrew achieved 69.0\% and 72.8\%, respectively, compared with 64.4\% and 66.9\% for Atomas. On the single-label endpoints, ranking analyses showed enrichment of positives among the highest-scoring candidates (Supplementary Table~\ref{tab:S5}). Absolute scores nevertheless varied across datasets: ClinTox and HIV scores were concentrated near zero despite high AUROC (Supplementary Table~\ref{tab:S9}; Supplementary Note~\ref{note:property-ranking}). These scores therefore support ranking within each endpoint, but should not be interpreted as calibrated probabilities of clinical outcomes.

We selected BACE, a dataset of human beta-secretase 1 inhibitors, to examine individual scores and retention of labelled positives after inspecting local score behaviour (Fig.~\ref{fig:geometry}c). At a threshold of 0.5, MolCrew retained 87.5\% of positives, compared with 68.8\% for a local logistic probe fitted to unchanged Atomas molecular representations. The two evaluation sets contained 152 and 151 examples, respectively, with 48 positives in each (Supplementary Table~\ref{tab:S8}). This local probe analysis is separate from the published Atomas comparison in the primary benchmark. The rationale for selecting BACE, including the smaller BBBP difference observed with the local probe, is detailed in Supplementary Note~\ref{note:geometry-construction}. Matched examples illustrate both higher-scoring positives and an active molecule assigned a lower score, showing useful ranking alongside a remaining failure (Supplementary Fig.~\ref{fig:si_case_property}).

\subsection{Instruction-directed molecular editing}\label{sec:edit}

We tested whether MolCrew could edit molecular structures to increase hydrogen-bond donor count (HBD), hydrogen-bond acceptor count (HBA) or drug-likeness (QED). Each condition used 500 DrugAssist starting molecules and the published two-round retrieval-feedback protocol~\cite{ye2025drugassist,bickerton2012qed}. Training included 1{,}500 loose-prompt pairs derived from benchmark test molecules (Methods; Supplementary Note~\ref{note:training-decoding}). Under this training configuration, success rates exceeded published DrugAssist results for loose HBA increases (86.0\% versus 71.0\%), loose HBD increases (85.0\% versus 72.0\%) and strict HBA increases (74.0\% versus 67.0\%; Fig.~\ref{fig:main}d; Supplementary Table~\ref{tab:S4}). QED success rates were similar under both loose (75.4\% versus 76.0\%) and strict (62.0\% versus 63.0\%) thresholds. For strict HBD increases, however, success was lower (68.4\% versus 76.0\%).

Retrieval feedback improved success relative to single-shot generation with the same editing checkpoint. In the original ablation, loose HBA success increased from 53.0\% to 86.0\%, and strict HBA success increased from 27.2\% to 74.0\% (Supplementary Fig.~\ref{fig:si_edit_ablation}; Supplementary Note~\ref{note:training-decoding}). HBD success also increased, whereas gains for QED were smaller. A separate comparison of training configurations showed that data augmentation increased loose HBD success from 55.6\% to 85.0\% and loose HBA success from 68.0\% to 86.0\% under the same retrieval-feedback procedure. These gains did not extend to every condition: strict QED success was 64.0\% with the base training pack and 62.0\% with the augmented pack.

\begin{figure}[pt]
\centering
\includegraphics[width=\linewidth]{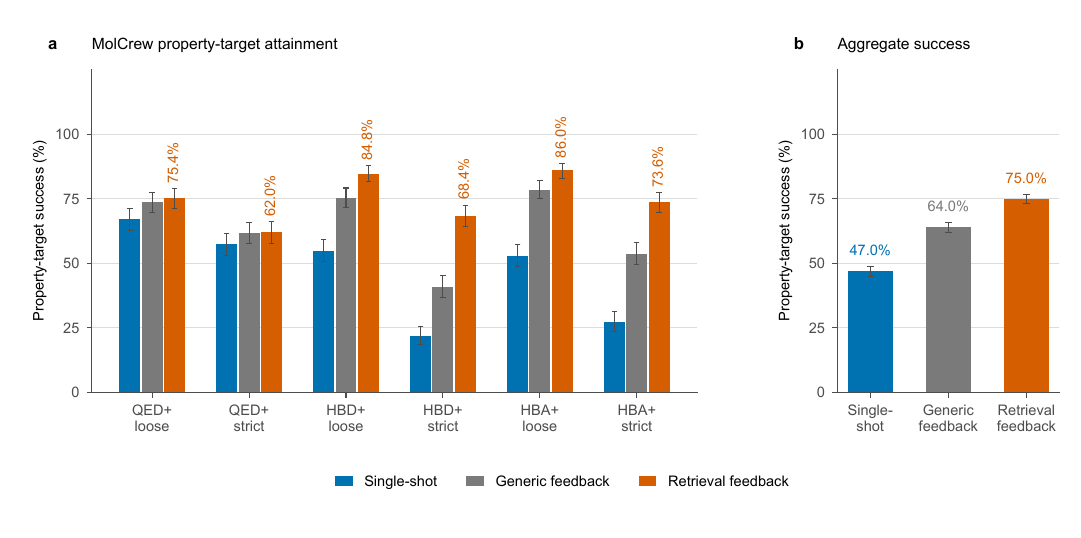}
\caption{\textbf{Feedback improves success in reaching molecular property targets.}
The same MolCrew editing checkpoint was evaluated using single-shot generation, generic feedback and retrieval feedback. Each strategy used 500 DrugAssist starting molecules across six property and threshold conditions, yielding 3{,}000 requests per strategy.
\textbf{a},~Success rates under the DrugAssist criterion for each condition, with 500 requests per group. Orange labels indicate retrieval-feedback rates. QED, quantitative estimate of drug-likeness; HBD, hydrogen-bond donor count; HBA, hydrogen-bond acceptor count. Loose and strict thresholds follow the DrugAssist protocol.
\textbf{b},~Overall success across all six conditions: 47.0\% (1{,}409/3{,}000) for single-shot generation, 64.0\% (1{,}920/3{,}000) for generic feedback and 75.0\% (2{,}251/3{,}000) for retrieval feedback.
Whiskers show descriptive 95\% percentile intervals from 10{,}000 bootstrap resamples clustered by starting molecule (seed~42). All six conditions and inference settings remain paired within each sampled molecule. Invalid and unsuccessful requests remain in the denominators. Training included pairs derived from benchmark test molecules, as detailed in Methods. These results describe attainment of numerical property targets in fixed runs of one specialist. They do not establish generalization to an untouched test set, exact scaffold retention, superiority at matched computational cost or a ranking of all available workflows.}
\label{fig:molcrew-feedback-followup}
\end{figure}

In a separate follow-up evaluation, we compared single-shot generation, generic feedback and retrieval feedback using the same MolCrew editing checkpoint. Each strategy was evaluated on the same 500 starting molecules across six DrugAssist property and threshold conditions, yielding 3{,}000 requests per strategy. Overall success in reaching the requested property target was 47.0\% (1{,}409/3{,}000) with single-shot generation, 64.0\% (1{,}920/3{,}000) with generic feedback and 75.0\% (2{,}251/3{,}000) with retrieval feedback (Fig.~\ref{fig:molcrew-feedback-followup}). For strict HBD increases, the corresponding rates were 22.0\%, 40.8\% and 68.4\%; for strict HBA increases, they were 27.4\%, 53.8\% and 73.6\%. These results show that retrieval feedback improved attainment of descriptor targets under this protocol. They do not establish exact scaffold retention or an advantage at matched computational cost.

We also compared individual loose HBA edits with locally available DrugAssist retrieval-feedback outputs (Fig.~\ref{fig:geometry}d). On the original 500 requests, MolCrew succeeded in 430 cases (86.0\%), compared with 336 (67.2\%) for DrugAssist. This local DrugAssist rate differs from the published 71.0\% used in the primary benchmark comparison. The matched subset included 16 cases in which DrugAssist succeeded while MolCrew left HBA unchanged. Four examples favouring MolCrew are shown in Supplementary Fig.~\ref{fig:si_case_edit}. Selection procedures, matching rules, figure exclusions and success categories are detailed in Supplementary Notes~\ref{note:case-selection} and~\ref{note:geometry-construction}. Success here refers to the requested descriptor change; additional structural constraints require separate checks.

\subsection{Coordinating molecular tasks}\label{sec:case}

We next evaluated how MolCrew combined molecular specialists and chemistry tools to address user requests. In this evaluation, the pretrained Qwen orchestrator operated without an adapter, selecting specialist or RDKit calls and using their outputs to guide subsequent steps. Prompt instructions governed tool selection, output checks and stopping behaviour (Fig.~\ref{fig:overview}; Supplementary Note~\ref{note:prompt-contracts}). Requests for descriptors alone could be handled directly by RDKit. Performance was assessed against the checks attached to each request, separately from the specialist benchmarks described above.

Across 1{,}699 recorded requests in 11 tiers, 1{,}554 (91.5\%) passed every check. Mean partial credit was 98.4\%, and 89.5\% of the 826 highest-priority requests passed all checks (Supplementary Figs~\ref{fig:si_orch} and~\ref{fig:si_serving_tiers}). The evaluation recorded 14{,}667 RDKit checks and 2{,}844 specialist calls. All 232 requests handled through the RDKit-only route and all 137 handled through the captioning route passed their checks (Supplementary Table~\ref{tab:S6}). These rates describe performance against recorded checks in one evaluation. They do not independently establish the correctness of every response or demonstrate an advantage over a simpler workflow.

Targeted property edits accounted for 90 of the 145 failed requests. Of these, 71 left the requested descriptor unchanged, six changed it without meeting the signed target, and 13 lacked a recorded value for the change (Supplementary Note~\ref{note:serving-residuals}). Image-upload requests passed in 43 of 60 cases (71.7\%), with failures concentrated in action selection or routing. This result measures request handling, not optical structure-recognition accuracy. The records also contained 189 empty-reply flags and two runtime errors (Supplementary Note~\ref{note:serving-ops}); loop counts are reported in Supplementary Table~\ref{tab:S7}. These failures highlight limitations in both reaching descriptor targets and completing the requested interaction.

Prompt refinement adjusted tool selection and recovery instructions while leaving Qwen and specialist weights unchanged. Recorded screening pass rates were 30.0\% for the initial prompt and 66.0\% and 84.0\% for two accepted early revisions. A later, expanded screen of 210 cases had a pass rate of 83.8\% (Supplementary Note~\ref{note:prompt-families}). The deployed instructions covered direct RDKit requests, retries for signed descriptor changes, scaffold constraints and responses when no further calls were available. These development screens were not a paired comparison on the 1{,}699-request evaluation and therefore do not quantify improvement on that set. The development history and verbatim prompts are provided in Supplementary Notes~\ref{note:prompt-families} and~\ref{note:prompt-contracts}.

\begin{figure}[pt]
\centering
\includegraphics[width=\linewidth]{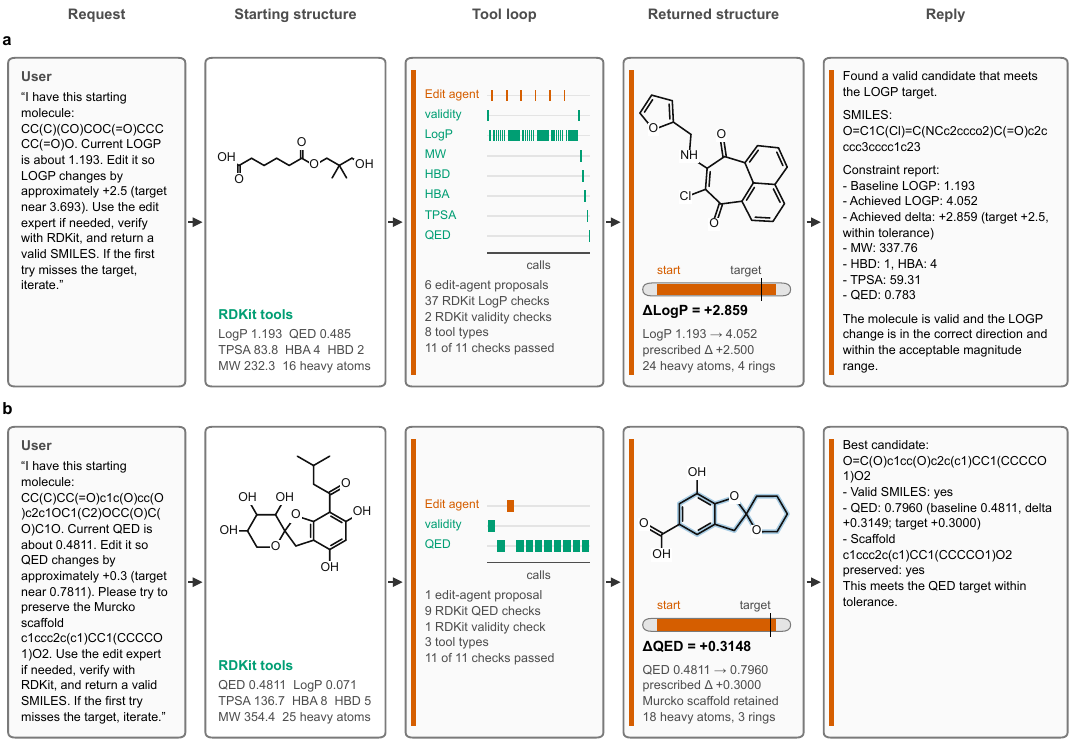}
\caption{\textbf{Two recorded molecular editing workflows.}
Each row traces a request from the starting structure through specialist and RDKit calls to the returned molecule and response. Descriptors were recomputed from the recorded structures using RDKit during figure preparation.
\textbf{a},~Editing to increase Crippen LogP by 2.500 without a scaffold constraint. LogP increased from 1.193 to 4.052, and the acyclic starting molecule was replaced by a product with four rings.
\textbf{b},~Editing to increase QED by 0.3000 while retaining a specified three-ring Bemis--Murcko scaffold. QED increased from 0.4811 to 0.7960, with the requested scaffold retained (shaded). Orange spans connect the starting and returned values; black ticks mark the requested final values.
Both examples were selected from 140 transcripts that passed every signed-shift check. Row \textbf{a} had the most iterations, and row \textbf{b} had the most rings in its specified scaffold among the scaffold-preserving QED cases. These selected examples do not estimate overall success rates. Selection rules and verbatim records are provided in Supplementary Notes~\ref{note:serving-transcripts} and~\ref{note:serving-verbatim}.}
\label{fig:case}
\end{figure}

Two recorded optimisation episodes illustrate how molecular proposals and descriptor checks were combined (Fig.~\ref{fig:case}). Both were selected from 140 transcripts that passed every signed-shift check: the episode with the most iterations and the scaffold-preserving QED case whose specified scaffold contained the most rings (Supplementary Note~\ref{note:serving-transcripts}). In the first episode, Crippen LogP increased from 1.193 to 4.052 against a requested increase of 2.500. The workflow used 15 iterations and 50 tool calls over 397.31~s. With no scaffold constraint, the edit changed an acyclic starting molecule into a product with four rings. In the second episode, QED increased from 0.4811 to 0.7960 against a requested increase of 0.3000, while retaining the specified three-ring scaffold. This workflow used five iterations and eleven tool calls over 93.31~s. These selected examples illustrate how the system completed requests with different structural constraints; they do not estimate the frequency of success. Verbatim requests, replies and checks are provided in Supplementary Note~\ref{note:serving-verbatim}.

\subsection{Linking tool-use decisions to molecular editing outcomes}\label{sec:orchestrator-adaptation}\label{sec:candidate-handling}

We examined whether improvements in tool-use decisions translated into successful molecular editing. In separate synthetic evaluations, the initial tool-use adapter increased exact agreement with prescribed next decisions from 31.3\% to 81.3\%, while the recovery adapter increased agreement from 37.5\% to 66.3\% on failure-and-recovery contexts. These gains measured the ability to reproduce specified decisions from supplied contexts. In a matched live comparison involving 60 editing requests that required both a descriptor target and retention of a specified molecular core, chemical success was unchanged with the recovery adapter, and automated workflow completion was lower. Improved agreement on synthetic decisions therefore did not establish improved completion of molecular tasks (Supplementary Tables~\ref{tab:S10} and~\ref{tab:S11}).

Candidate handling provided a separate test of whether MolCrew could deliver suitable structures generated by its editing specialist. We compared candidate-ranking and acceptance procedures while keeping the model weights, recovery adapter, prompts, seeds and permitted budgets unchanged. The comparison used 36 previously unexposed evaluation cases from the ChEBI training partition, with requests to increase HBD, HBA or QED. Success required a valid changed structure and the requested descriptor increase, without the core-retention requirement of the adaptation pilot. Prioritising candidates that satisfied the descriptor target and allowing verified editor candidates to be retained in the final answer increased editing success from 50.0\% (18/36) to 86.1\% (31/36), with 13 paired gains and no observed regressions. A fixed workflow using the same editor and RDKit achieved 91.7\% (33/36).

Inspection of the paired outputs showed that 12 of the 13 gains recovered successful structures already present in the same initial candidate pool. For these cases, the editor had generated a suitable molecule, but candidate handling prevented its delivery in the original workflow. The improvement therefore demonstrates a contribution from the software that connects specialist generation to the final answer, without additional model training. This single-seed comparison supports more reliable execution of the tested descriptor-editing tasks; it does not establish improved core preservation or an advantage over the fixed workflow (Supplementary Tables~\ref{tab:S10} and~\ref{tab:S11}).

\section{Discussion}\label{sec:discussion}

MolCrew provides a modular approach to molecular design in which separately trained specialists are coordinated through explicit tool calls and chemical checks. This organisation reflects the linked decisions of early medicinal chemistry: proposing structures, assessing their properties and choosing the next modification~\cite{hughes2011principles,paul2010improving,vamathevan2019applications}. Unified molecule--language models bring several capabilities into shared representations or model parameters~\cite{luo2023molfm,zhang2025atomas,edwards2022molt5,christofidellis2023unifying,liu2023moleculestm,liu2024gitmol}. In MolCrew, separate specialist weights allow each capability to be evaluated and updated independently. This design introduces a coordination burden, including tool interfaces, prompt instructions and output checks, but makes the contribution of each component easier to inspect. The study evaluates this design choice without establishing that modularity is necessary for accurate or auditable molecular workflows.

The specialist results show both the strengths and the limits of this approach. Structure generation achieved higher exact recovery than published comparators, while molecular descriptions showed closer agreement with reference text under the reported scoring protocol~\cite{edwards2022molt5,zhang2025atomas,beltagy2019scibert}. Mean AUROC exceeded published Atomas results across all eight MoleculeNet datasets evaluated using scaffold splits~\cite{wu2018moleculenet}. Editing performance depended more strongly on the objective: success exceeded published DrugAssist results for several hydrogen-bond count conditions, whereas QED results were similar and strict donor-count success was lower~\cite{ye2025drugassist,bickerton2012qed}. Comparisons using the same editing checkpoint also showed that feedback improved attainment of numerical property targets. Fingerprint analyses and matched examples reveal remaining structure mismatches, missed positives and unsuccessful edits (Fig.~\ref{fig:geometry}; Supplementary Figs~\ref{fig:si_case_textmol}--\ref{fig:si_case_edit} and~\ref{fig:si_mismatch}). These differences help define where each specialist may be useful and where further validation is needed.

The recorded workflow evaluation examined how these capabilities operated together. Using unadapted, prompt-guided Qwen, 1{,}554 of 1{,}699 requests (91.5\%) passed every recorded check (Supplementary Note~\ref{note:eval-inventory}). This is an operational measure of compliance with the attached checks, not an independently verified measure of response correctness. Most failed requests involved targeted property edits, often because the requested descriptor remained unchanged; image-request errors and empty-reply flags identified additional limitations in request handling (Supplementary Notes~\ref{note:serving-residuals} and~\ref{note:serving-ops}). Prompt refinement changed tool-selection and recovery instructions while keeping Qwen and specialist weights fixed (Fig.~\ref{fig:prompt_opt}). However, the development screens were not a paired comparison on the final request set, so they do not quantify the improvement attributable to prompt refinement. The selected LogP and scaffold-preserving QED examples illustrate how proposals and calculations can be combined and inspected, but do not estimate typical success rates or computational cost (Fig.~\ref{fig:case}).

The matched evaluations distinguish choosing a tool action from carrying a suitable molecular proposal through to the final response. Most gains from revised candidate handling recovered structures already generated by the editing specialist. Failures can therefore arise after successful generation, making candidate ranking and acceptance part of molecular workflow performance. These improvements required no additional model training. Orchestrator adaptation addressed a different aspect of coordination: agreement with prescribed synthetic decisions improved, but chemical success was unchanged and automated workflow completion was lower in the separate core-constrained pilot. The findings support evaluating learned decisions through their executed molecular outcomes, alongside agreement with prescribed actions. The strong performance of the fixed editor--RDKit workflow also provides an important reference for assessing what learned orchestration contributes to these bounded editing tasks.

Several features of the evaluation limit the conclusions that can be drawn. The specialist benchmarks and matched workflow studies used fixed checkpoints with decoding settings specified for each evaluation. The matched live comparisons used one generation seed and different editing criteria: the adaptation pilot required retention of a specified molecular core, whereas the candidate-handling study evaluated descriptor-directed editing without that requirement. The candidate-handling cohort was drawn from the ChEBI training partition, so its results do not establish generalization to molecules unseen during specialist training. The descriptive bootstrap intervals in the editing-feedback analysis capture variation across sampled benchmark molecules, not variation across retraining or repeated generation. Editing training included pairs derived from DrugAssist test molecules, so those results do not establish generalization to an untouched test set (Supplementary Table~\ref{tab:S4}). Primary comparisons use published baseline results, whereas local diagnostic analyses use available outputs or probes. For example, caption diagnostics use MolT5-large because suitable Atomas captions were unavailable, while the primary comparison retains published SciBERT-token scores (Supplementary Table~\ref{tab:S2}). Illustrative examples were selected, often with size restrictions for legibility, and should not be treated as representative samples (Supplementary Note~\ref{note:case-selection}). Image-upload outcomes likewise describe request handling rather than optical structure-recognition accuracy.

The evidence also remains computational. Molecular representations and RDKit checks support structural validity and the assessment of specified descriptor constraints~\cite{krenn2020selfies,landrum2013rdkit}; they do not establish biological activity or experimental suitability. Property scores support ranking within individual endpoints and should not be interpreted as calibrated probabilities of clinical outcomes or substitutes for experimental assessment of absorption, distribution, metabolism, excretion and toxicity. No docking or target-binding performance was evaluated. Further work should therefore test the system on untouched molecules, compare coordination strategies under matched computational budgets and assess whether the resulting candidate rankings and edits are useful experimentally.

Local deployment offers a practical setting for this work. Language models provide a flexible interface for expressing requests~\cite{brown2020language}, while locally served specialists allow molecular workflows to remain within an organisation's data environment~\cite{price2019privacy}. Recorded prompts, tool calls and outputs support inspection of how a result was obtained, and separate model weights allow capabilities to be maintained independently (Supplementary Note~\ref{note:prompt-contracts}). This approach complements broader scientific agent systems that coordinate specialized components across discovery tasks~\cite{ghareeb2026robin}. Retrosynthesis, reaction prediction, multi-objective editing and interpretation of molecular sketches are possible extensions~\cite{schwaller2019molecular,gomez2018automatic,wang2025sketchmol}, each requiring its own validation and assessment within the combined workflow. MolCrew thus provides a framework for connecting molecular capabilities while keeping their outputs, evaluation criteria and limitations visible as candidates are considered for experimental investigation.

\section{Conclusion}\label{sec:conclusion}

MolCrew brings together four molecular specialists in a locally deployable orchestration harness for structure generation, molecular description, endpoint ranking and instruction-directed editing. RDKit checks link molecular proposals to explicit calculations and structural constraints. On ChEBI-20, exact structure recovery reached 96.4\%, compared with 38.7\% for published Atomas-large results. Across eight MoleculeNet datasets evaluated using scaffold splits, mean AUROC was 85.1\%, compared with 77.0\% for published Atomas results. In a recorded evaluation using the unadapted orchestrator, 91.5\% of 1{,}699 requests passed every predefined check. These results describe performance under the evaluated protocols, with limitations that include fixed checkpoints and overlap between editing training data and benchmark test molecules. They do not establish docking performance or experimental ADMET properties. The practical contribution is a modular framework in which specialists can be evaluated independently and molecular proposals remain linked to recorded tool outputs and chemical checks. This design supports computational candidate assessment and refinement while allowing organisations to retain local control of molecular data.

\section{Methods}\label{sec:methods}

\methodheading{Shared backbone}
The four specialists share a chemistry-aware decoder-only transformer $f_\theta$~\cite{vaswani2017attention,raffel2020exploring} but are trained, versioned, and served as separate models with task-specific heads and decoding settings. Molecules are tokenised as SELFIES sequences $y=(y_1,\ldots,y_T)$, which map every string to a chemically valid graph~\cite{krenn2020selfies}; text uses a hybrid subword vocabulary. User-facing structures are returned as SMILES after RDKit sanitisation~\cite{weininger1988smiles,landrum2013rdkit}. Writing $\mathrm{Decode}(y)$ for SELFIES-to-graph decoding and $\mathrm{SMILES}(\cdot)$ for the RDKit canonical string, generation is the composition $x\mapsto y\mapsto \mathrm{Decode}(y)\mapsto \mathrm{SMILES}(\mathrm{Decode}(y))$. Sanitisation failure is counted as Validity $=0$. Checkpoint identifiers and decode settings are archived in Source Data (Note~\ref{note:eval-inventory}). Training uses AdamW with a cosine learning-rate schedule and gradient clipping; optimiser hyperparameters for each reported checkpoint are archived with Source Data. No specialist is fine-tuned on the ChEBI-20, MoleculeNet scaffold-test, or DrugAssist evaluation splits, except for the disclosed DrugAssist loose-prompt overlap noted under editing.

Generative specialists minimise next-token cross-entropy. For context $x$ and target tokens $y=(y_1,\ldots,y_T)$,
\begin{equation}
\mathcal{L}_{\mathrm{CE}}(\theta;x,y)=-\sum_{t=1}^{T}\log p_\theta\!\left(y_t\mid y_{<t},x\right),
\end{equation}
with $p_\theta(y_t\mid y_{<t},x)=\mathrm{softmax}\!\left(W h_t(x,y_{<t})\right)$ the categorical distribution over the task vocabulary at hidden state $h_t$. Mini-batches average $\mathcal{L}_{\mathrm{CE}}$ over sequences. No reinforcement learning, preference ranking, or docking-derived reward is used. Property heads are trained with the explicit binary cross-entropy below rather than with a fused logits criterion that hides the sigmoid.

\methodheading{Orchestrator adaptation}

We trained independent tool-use and recovery adapters from the same unadapted Qwen3.8-27B checkpoint, revision \texttt{1d4bf0f2ff6012fd82039f2fa52739d0dd7c60c0}, while leaving the base-model parameters and all specialist weights unchanged. The tool-use study used 510 training, 16 validation and 32 test examples; the recovery study used 512 training examples and separate validation and test sets of 80 examples each. Examples paired a supplied context with a prescribed next decision. Recovery contexts combined calculated molecular observations with synthetic tool failures and specified specialist outputs; they were not complete trajectories generated by executing the specialists.

Both adapters used low-rank adaptation~\cite{hu2022lora} with rank 8, scaling parameter 16 and zero dropout on 304 text-attention projection modules. We minimised causal token cross-entropy over the target assistant completion, including the target JSON reply and end-of-sequence token but excluding prompt tokens. Training used one epoch, a batch size of one, gradient accumulation over up to eight examples, AdamW with a constant learning rate of $2\times10^{-5}$, zero weight decay, gradient-norm clipping at 1.0 and seed 20261006. Base-model parameters used BF16 precision; trainable adapter parameters used FP32 with BF16 autocast. Maximum sequence lengths were 8{,}448 and 9{,}728 tokens for the tool-use and recovery studies, respectively. Each run completed 64 optimiser updates, and we evaluated the final adapter without selecting a checkpoint on validation or test performance.

Synthetic decision evaluation used generation without sampling and a maximum of 640 new tokens. Exact decision agreement required a response that passed the study's schema checks and matched the prescribed action, route, ordered tool calls and arguments, and SMILES string for the working molecule. Exact matching of the narrative reply was not required.

\methodheading{Matched live evaluation of orchestrator adaptation}

We compared unadapted Qwen with the recovery adapter using the same execution software, specialists, requests and permitted budgets. Inputs were drawn from the ChEBI training partition. Each condition received 24 development requests for execution readiness and 80 separate final requests: 20 each for increased HBD, increased HBA, increased QED and measurement of the current molecule. HBD and HBA targets required an increase of at least one; QED targets were the initial value plus 0.02, rounded to four decimal places. Before inference, rule-based candidate structures were used to check that the requested targets and core constraints were attainable; these structures were withheld from the models. No artificial tool failures were injected into the live trajectories.

Chemical editing success required a valid, connected structure that differed from the input, had been returned by the editor, met the descriptor target and retained the specified input Bemis--Murcko core as an atom-and-bond subgraph~\cite{bemis1996scaffold}. Graph identity and core matching did not require stereochemical agreement. Automated workflow success additionally required valid molecular lineage through the editor calls, compliance with the editor budget, and a correct descriptor measurement for the returned graph displayed in the response. Current-molecule requests required the supplied graph to remain unchanged, with molecular weight, Crippen LogP and TPSA correctly calculated and displayed without editing.

\methodheading{Candidate-handling comparison}

We compared candidate ranking and acceptance before and after the software revision, alongside a fixed editor--RDKit workflow. The revision ranked candidates satisfying all supplied descriptor constraints ahead of candidates failing strict boundaries and allowed a generated candidate to be accepted after verifying its provenance and candidate-specific constraints. Both MolCrew conditions used the same specialists, recovery adapter, prompts, generation seeds and permitted budgets. Each condition received eight readiness requests and 48 final requests, with 12 final requests for each of HBD increase, HBA increase, QED increase and current-molecule measurement. Inputs came from the ChEBI training partition after excluding recorded public holdouts, earlier exposed molecular identities, cores and source identifiers, and development fixtures. These exclusions did not establish absence from specialist training.

The fixed workflow measured the original descriptor and requested up to two editor batches of eight candidate attempts each. It selected the valid candidate with the highest requested descriptor value and stopped when that value exceeded the original baseline. If the first batch was unsuccessful, fixed feedback supplied the original and best observed values before another edit of the original input.

Across all three conditions, chemical success required a nonempty valid changed graph, verified editor provenance and a strictly increased requested descriptor. Structural similarity, core retention and connectivity were offline diagnostics rather than success requirements; QED success did not require the 0.02 increase used in the adaptation pilot. Similarity used Morgan fingerprints of radius 2 and 2{,}048 bits with Tanimoto comparison. Automated workflow success additionally required the requested final-graph measurement, its correct display and compliance with the declared budgets. Shared software completed final-molecule measurements across conditions, so this endpoint assessed the delivered output rather than whether the learned planner independently requested every calculation. The study limited each request to 600~s, 64 explicit RDKit calls and three final-measurement calls; descriptor calculations internal to editor ranking were excluded from the explicit-call limit.

\methodheading{Execution and scoring of matched workflows}

Both matched studies used BF16 Qwen inference and FP32 molecular specialists. Qwen conditions allowed six planning turns, at most 640 new tokens per decision, temperature 0.3, top-$p$ 0.9 and a total context capacity of 65{,}536 tokens. Editing requests permitted at most two editor calls and eight candidate attempts per call. Each editor attempt allowed up to 256 new tokens within the specialist's context capacity. The first candidate was generated greedily; subsequent attempts used temperatures 0.7, 0.9, 1.0, 1.1, 1.2, 1.3 and 1.4. Random states were reset for each request using the same case-derived seed across matched conditions. Final cohorts and scoring criteria were fixed before inference, and readiness cases were not used for outcome-based tuning.

We scored saved outputs using RDKit 2024.03.2 and independently checked final descriptors and recorded editor provenance. Every assigned request remained in the denominator, including failed and missing outputs. Editing and current-molecule measurement requests were analysed separately. We recorded paired gains and regressions and calculated descriptive 95\% percentile intervals from 10{,}000 bootstrap resamples of paired requests. These intervals quantify variation across requests under one matched generation seed, not variation across retraining or repeated generation. Actual calls, tokens and elapsed time were recorded separately; matching permitted budgets did not imply equal computational cost. Automated checks assessed the specified molecular and reporting criteria, not unrestricted factual accuracy of the narrative response.

\methodheading{Original orchestration and prompt refinement}

Fig.~\ref{fig:overview} shows the serving layout used in the original recorded evaluation, in which the pretrained Qwen orchestrator operated without an adapter. Tool selection and routing were governed by prompt instructions and observed tool outputs; no separate routing classifier was trained for this evaluation. Let $\mathcal{T}$ be the allowlisted tool set. Write $H_0$ for the initial context (the user request together with the active prompt contract $P$). A request $q$ is resolved in at most $L=15$ turns. At turn $\ell$ the orchestrator emits a plan $(a_\ell,r_\ell,C_\ell)$, where $a_\ell$ is the recorded action, $r_\ell$ the resolved route, and $C_\ell\subset\mathcal{T}$ the tool batch. Each tool $\tau\in C_\ell$ returns an observation $o_{\ell,\tau}$, and the context is updated by
\begin{equation}
H_{\ell}=H_{\ell-1}\cup\bigl\{(a_\ell,r_\ell,C_\ell)\bigr\}\cup\{o_{\ell,\tau}:\tau\in C_\ell\}.
\end{equation}
The episode ends when $a_\ell$ is a terminal answer, the loop budget is exhausted ($\ell=L$), or a refuse / reject-upload action is issued. A request passes if and only if every recorded chemistry check returns true. No parameters of a task-routing classifier are trained; $r_\ell$ is part of the harness plan under $P$.

\begin{figure}[pt]
\centering
\includegraphics[width=0.48\textwidth]{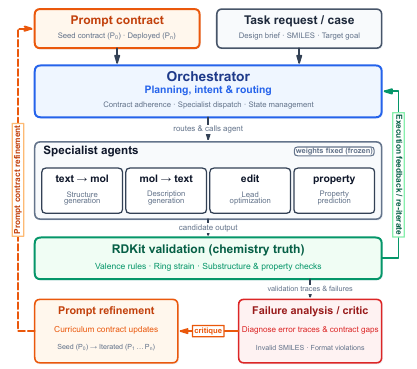}
\caption{\textbf{Prompt-contract refinement with frozen specialists.}
Task requests and contracts feed the orchestrator, which plans tools and routes to \TextMol{}, \MolText{}, edit, or property agents (weights fixed). RDKit supplies calculated descriptors and checks of specified structural constraints. Inner loop: in-context correction on failures. Outer loop: critic edits contracts from $P_0$ to $P_n$. Reported routing is harness planning, not a trained router. The reported serving pass rate uses the deployed contract $P_n$ (Note~\ref{note:eval-inventory}).}
\label{fig:prompt_opt}
\end{figure}

Candidate structures are checked with RDKit before return; failed checks can trigger in-context correction within the loop budget. Serving behaviour is governed by prompt contracts: text that specifies allowed tools, what counts as a valid answer, and how to recover from chemical failures. The deployed prompts are reproduced in Note~\ref{note:prompt-contracts}. Contracts were refined with specialist weights frozen (Fig.~\ref{fig:prompt_opt}). An outer critic proposes prompt edits from failure traces; an inner loop screens those edits against chemistry ground-truth checks, yielding a deployed contract $P_n$ from a seed contract $P_0$. Recorded screen decisions are summarised in Note~\ref{note:prompt-families}. The reported serving pass rate uses $P_n$ and is not a paired $P_0$ versus $P_n$ ablation on the same requests (Supplementary Fig.~\ref{fig:si_orch}; Supplementary Fig.~\ref{fig:si_serving_tiers}; Note~\ref{note:eval-inventory}).

\methodheading{Text-to-structure generation}
\TextMol{} minimises $\mathcal{L}_{\mathrm{CE}}$ with $x$ a natural-language description and $y$ the target SELFIES sequence, then converts $y$ to SMILES for scoring~\cite{krenn2020selfies,landrum2013rdkit}. Training uses public description--molecule corpora; the reported model uses an expanded public mix relative to a ChEBI+PCdes base (Supplementary Fig.~\ref{fig:si_textmol_ablation})~\cite{edwards2022molt5,hastings2016chebi,zhang2025atomas}. After decoding, the prediction is sanitised with RDKit. Writing $\mathrm{InChI}(\cdot)$ for the RDKit InChI of a sanitised graph,
\begin{equation}
\mathrm{Exact}(\hat{m},m^\star)=\mathbf{1}\!\left[\mathrm{InChI}(\hat{m})=\mathrm{InChI}(m^\star)\right],\qquad
\mathrm{Validity}(\hat{m})=\mathbf{1}\!\left[\mathrm{Sanitize}(\hat{m})=1\right].
\end{equation}
Fingerprint scores are Tanimoto coefficients on MACCS, RDKit, and Morgan bit vectors (radius 2, 2{,}048 bits for Morgan)~\cite{rogers2010extended}. Character-level string overlap is not substituted for Exact. Reported decoding uses beam size~1.

\methodheading{Structure-to-text description}
\MolText{} minimises $\mathcal{L}_{\mathrm{CE}}$ with roles reversed: $x$ is the molecular input and $y$ the reference caption tokens. Training uses reverse pairs from the same public caption family~\cite{edwards2022molt5,zhang2025atomas}. Reported decoding uses beam size~5. Caption scores are SciBERT-token BLEU-2/4 and ROUGE-1/2/L under the Atomas Table~3 WordPiece recipe (Supplementary Table~\ref{tab:S2})~\cite{papineni2002bleu,lin2004rouge,beltagy2019scibert}. Let $r$ and $\hat{c}$ be the SciBERT WordPiece sequences of the reference and hypothesis. BLEU-$n$ is clipped $n$-gram precision with the standard brevity penalty on those sequences; ROUGE-$n$ and ROUGE-L are the corresponding recall-oriented overlaps. Geometry and case-study panels use a different similarity family (shared-space TF-IDF word 1--2 grams versus the ChEBI-20 reference) and are not substituted for SciBERT BLEU. When a published Atomas mol$\to$text checkpoint is unavailable, local per-example geometry uses MolT5-large smiles2caption as a comparator, while Table~\ref{tab:S2} retains published Atomas-base SciBERT numbers.

\methodheading{Multi-task property prediction}
Eight MoleculeNet scaffold-classification heads $h\in H=\{\mathrm{BBBP},\mathrm{Tox21},\mathrm{ToxCast},\mathrm{SIDER},\mathrm{ClinTox},\mathrm{MUV},\mathrm{HIV},\mathrm{BACE}\}$ share a molecular representation~\cite{wu2018moleculenet}. Let $\sigma$ denote the logistic sigmoid. Binary cross-entropy between a logit $z\in\mathbb{R}$ and a label $\ell\in\{0,1\}$ is
\begin{equation}
\ell_{\mathrm{BCE}}(z,\ell)=-\bigl[\ell\log\sigma(z)+(1-\ell)\log\bigl(1-\sigma(z)\bigr)\bigr].
\end{equation}
For head $h$, write $\mathcal{A}_h$ for the assays with observed labels on that example, $z_{h,a}$ for the corresponding logit, and $\ell_{h,a}$ for the label. Single-task heads have $|\mathcal{A}_h|=1$; multi-label heads (Tox21, ToxCast, SIDER, MUV) contribute only on observed assays. The per-head loss is the mean over those assays,
\begin{equation}
\overline{\ell_{\mathrm{BCE}}}(z_h,\ell_h)=\frac{1}{|\mathcal{A}_h|}\sum_{a\in\mathcal{A}_h}\ell_{\mathrm{BCE}}(z_{h,a},\ell_{h,a}),
\end{equation}
and the batch objective averages equally over labelled heads $H_{\mathrm{lab}}$,
\begin{equation}
\mathcal{L}_{\mathrm{prop}}(\theta)=\frac{1}{|H_{\mathrm{lab}}|}\sum_{h\in H_{\mathrm{lab}}}\overline{\ell_{\mathrm{BCE}}}\!\left(z_h,\ell_h\right).
\end{equation}
Training uses random SMILES enumerations of the same sanitised graph as alternative string views. At inference, sigmoid scores are averaged over $V=4$ independent random SMILES views,
\begin{equation}
\hat{s}_h=\frac{1}{V}\sum_{v=1}^{V}\sigma\!\left(z_h^{(v)}\right),
\end{equation}
where $z_h^{(v)}$ is the head logit for view $v$. Logits remain unconstrained reals; probabilities enter only through $\sigma$ in $\ell_{\mathrm{BCE}}$ and in $\hat{s}_h$. Equivalent numerically stable evaluations of $\ell_{\mathrm{BCE}}$ are used in training provided they implement this same loss. Scaffold splits follow Bemis--Murcko scaffolds~\cite{bemis1996scaffold}. Per-endpoint and mean ROC-AUC are in Supplementary Table~\ref{tab:S3}; ranked-budget enrichment on single-label endpoints is in Supplementary Table~\ref{tab:S5}.

\methodheading{Instruction-directed editing}
The edit specialist minimises $\mathcal{L}_{\mathrm{CE}}$ with $x=(\mathrm{seed},\mathrm{instruction})$ and $y$ the product SELFIES, using DrugAssist MolOpt-Instructions-style pairs~\cite{ye2025drugassist}, including 1{,}500 loose-prompt pairs seeded from the DrugAssist test set (disclosed with Supplementary Table~\ref{tab:S4}). Reported decoding is greedy. Drug-likeness targets use QED~\cite{bickerton2012qed}; hydrogen-bond counts use RDKit Lipinski donors and acceptors~\cite{lipinski2001experimental,landrum2013rdkit}. Training-pack and single-shot versus retrieval-feedback ablations are in Supplementary Fig.~\ref{fig:si_edit_ablation}.

Correct rate follows the published DrugAssist retrieval-feedback (ReDF) protocol with a maximum of $R=2$ feedback rounds after the first generation (up to three decodes)~\cite{ye2025drugassist}. Write $s_0$ for the seed SMILES and $g$ for the signed instruction. Let $r_{<t}$ be empty at $t=0$ and otherwise the retrieved neighbours of $s^{(t-1)}$ together with the recorded failure reason. For $t=0,\ldots,R$,
\begin{equation}
s^{(t)}\sim p_\theta\!\left(\,\cdot\mid s_0,g,r_{<t}\right).
\end{equation}
The loop stops at the first $t$ for which the published success predicate $\mathrm{OK}(s_0,s^{(t)},g)$ holds, or after $t=R$; invalid SMILES skip further retrieval. The episode is correct if the returned product $s^\star$ satisfies $\mathrm{OK}(s_0,s^\star,g)$. For hydrogen-bond instructions, $\mathrm{OK}$ requires chemical validity and a signed change in the RDKit Lipinski donor or acceptor count that meets the loose or strict threshold in DrugAssist Table~5; for QED+ it requires a signed increase in QED under the same table. Validity for the companion free-generation column is $\mathbf{1}[\mathrm{Sanitize}(s)=1]$ on a single decode without ReDF. No extra reward model or docking score is introduced.

\methodheading{Follow-up comparison of editing feedback}
We compared single-shot generation, generic feedback and retrieval feedback using the same editing checkpoint and 500 DrugAssist starting molecules under six property and threshold conditions~\cite{ye2025drugassist}. Single-shot outcomes were taken from the saved initial predictions of the completed retrieval-feedback run. Generic feedback reused these exact initial predictions and allowed up to two additional generations. After an unsuccessful valid proposal, the feedback identified the proposed molecule as incorrect and requested a new SMILES satisfying the original requirements, without supplying a retrieved exemplar. Subsequent prompts retained the original request and concatenated the accumulated feedback.  Retrieval feedback additionally supplied one exemplar from the DrugAssist molecular database, which contained 169{,}249 unique SMILES. Eligible molecules had a stored descriptor value exceeding the original input value by the condition-specific threshold; the original input string was excluded. Among eligible molecules, the exemplar maximised Morgan-fingerprint Tanimoto similarity to the preceding unsuccessful proposal, using radius 2, 2{,}048 bits and no chirality.

Retrieval eligibility remained anchored to the original molecule throughout the episode. Requests stopped upon success, exhaustion of two feedback rounds, or a terminal parsing, validity or retrieval failure. The parser rejected exact copies of the retrieved exemplar, and outputs longer than 200 SMILES characters were counted as failures.

Generic-feedback generation used FP32 model parameters, greedy decoding, seed 42 and at most 256 new tokens within a 1{,}024-token sequence limit. Success was evaluated relative to the original molecule using the upstream DrugAssist predicate. Loose conditions required a positive descriptor change. Strict HBD and HBA conditions required an increase greater than one, and strict QED required an increase greater than 0.1; the upstream QED exception accepted any increase when the initial QED exceeded 0.9. Numerical success did not require exact scaffold retention or a minimum structural similarity. All assigned requests, including invalid and unsuccessful outputs, remained in the denominators. This was a post-hoc paired comparison; the strategies used different numbers of new generations and different access to retrieved exemplars.

\methodheading{Serving episode and pass predicate}
Let $q$ be a recorded request and $\mathcal{C}(q)$ the fixed checks attached to that request (action, route, required and forbidden tools, loop cap, SMILES validity, and any signed property constraints). An episode has at most $L=15$ turns as above. Write $c(q)\in\{0,1\}$ for whether check $c$ returns true on the recorded transcript. Define
\begin{equation}
\mathrm{Pass}(q)=\prod_{c\in\mathcal{C}(q)} c(q)\in\{0,1\}.
\end{equation}
The request passes if and only if $\mathrm{Pass}(q)=1$. Partial credit is $|\mathcal{C}(q)|^{-1}\sum_{c\in\mathcal{C}(q)}c(q)$. On the recorded serving corpus $\mathcal{Q}$ (Note~\ref{note:eval-inventory}), the reported pass rate is $|\mathcal{Q}|^{-1}\sum_{q\in\mathcal{Q}}\mathrm{Pass}(q)$. No request is dropped for runtime error when computing that rate; recorded runtime errors fail the no-runtime-error check. Routing is the recorded $r_\ell$, not a learned classifier.

The allowlisted set is $\mathcal{T}=\mathcal{T}_{\mathrm{spec}}\cup\mathcal{T}_{\mathrm{rdkit}}\cup\mathcal{T}_{\mathrm{img}}$: four specialists, eight RDKit descriptor tools, and one image-recognition backend used only in the recorded evaluation (Note~\ref{note:prompt-contracts}). Plans with $C_\ell\not\subset\mathcal{T}$ fail the forbidden-tool check. RDKit-only resolutions correspond to $C_\ell\subset\mathcal{T}_{\mathrm{rdkit}}$ with no specialist route; they pass throughout the recorded serving corpus (Supplementary Table~\ref{tab:S6}). Shell, Python execution, and unrestricted code evaluation are forbidden. Specialist agents are invoked as tools with structured arguments; they do not carry a separate user-facing system prompt in the reported serving setup.

Signed property constraints are checked with RDKit on the recorded seed $s_0$ and product $s^\star$. For a request that asks for a signed shift $\Delta^\star$ in a descriptor $d$ (for example Crippen LogP or QED), write $\Delta=d(s^\star)-d(s_0)$. When $\Delta^\star\neq 0$, the direction check requires $\mathrm{sign}(\Delta)=\mathrm{sign}(\Delta^\star)$; a recorded $\Delta=0$ fails. Magnitude checks, when present in $\mathcal{C}(q)$, require $|\Delta-\Delta^\star|$ to fall inside the recorded tolerance. Scaffold-preserving requests additionally require that the Bemis--Murcko scaffold of $s^\star$ match the named scaffold string after RDKit canonicalisation~\cite{bemis1996scaffold}. These predicates are the chemistry ground truth behind Fig.~\ref{fig:case} and Supplementary Fig.~\ref{fig:si_orch}; they are not docking or target-binding scores.

\methodheading{Prompt-contract refinement}
A prompt contract $P$ specifies allowed tools, valid-answer rules, and recovery from chemical failures. Specialist weights $\theta$ are frozen while $P$ is edited. An outer critic proposes a candidate $P'$ from failure traces of a screen $\mathcal{Q}_{\mathrm{scr}}$. An inner loop serves $\mathcal{Q}_{\mathrm{scr}}$ under $P'$ and accepts $P'$ only if the recorded screen decision is an accept (Note~\ref{note:prompt-families}). Only accepted contracts enter the lineage $P_0\to\cdots\to P_n$. The deployed contract $P_n$ is the serving prompt used in the reported evaluation (22{,}072 characters). The reported pass rate uses $P_n$ on $\mathcal{Q}$ and is not a paired $P_0$ versus $P_n$ ablation on the same requests. Seed and deployed prompts are reproduced in Note~\ref{note:prompt-contracts}. Character counts of the stored lineage (seed 918; loop-evolved intermediate 2{,}576; deployed 22{,}072) summarise how much instruction text changed while $\theta$ stayed fixed (Fig.~\ref{fig:prompt_opt}; Note~\ref{note:prompt-families}).

\methodheading{SI case-study selection and depiction}
Supplementary Figs~\ref{fig:si_case_textmol}--\ref{fig:si_case_edit} are drawn from the same frozen evaluation records as Tables~\ref{tab:S1}--\ref{tab:S4}. Rules were fixed before inspection of structures: for \TextMol{} and property, two MolCrew-favouring cases, one near-tie or both-correct case, and one hard case; for edit, four MolCrew-only HBA+ successes versus DrugAssist ReDF; for \MolText{}, three drawable panels. Selection is deterministic (first / spaced / last index in the filtered pool). \TextMol{}, \MolText{}, and edit galleries restrict to 10--32 heavy atoms so 2D depictions remain legible; BACE actives are often larger and are not size-filtered (heavy-atom counts 49, 42, 16, and 31; Note~\ref{note:case-selection}). Indices: \TextMol{} $0$, $3298$, $1$, $936$; \MolText{} $4$, $3299$, $36$, $22$; property $92$, $2$, $27$, $8$. The galleries are not substitutes for the primary evaluations. Stored structures, captions, and scores appear with those figures.

Structures are RDKit 2D depictions. Coordinates use RDKit CoordGen when available, otherwise the default RDKit 2D layout. Bonds and heteroatom labels are stroked as native vector paths (the same family as Fig.~\ref{fig:overview} and Fig.~\ref{fig:case}), with Arial / Helvetica, PDF font type 42, and no SMILES string as the primary visual. Geometry-panel Morgan Tanimoto uses radius-2 bit vectors of length 2{,}048 as a drawable proxy; Exact on the structure evaluation remains InChI string equality (Note~\ref{note:board-metrics}).

\methodheading{Evaluation protocols}
\TextMol{} and \MolText{} are scored on the ChEBI-20 test split under Atomas-compatible protocols (Fig.~\ref{fig:main}a,b)~\cite{edwards2022molt5,zhang2025atomas}. Structure metrics are InChI exact match, MACCS/RDKit/Morgan Tanimoto, and RDKit validity~\cite{rogers2010extended,landrum2013rdkit}. Caption metrics are SciBERT-token BLEU and ROUGE~\cite{beltagy2019scibert,papineni2002bleu,lin2004rouge}. SciBERT WordPiece tokenisation follows the Atomas Table~3 recipe; Fig.~\ref{fig:geometry}b uses a different family (shared-space TF-IDF word 1--2 grams) and is not substituted for SciBERT BLEU. Property evaluation is MoleculeNet scaffold ROC-AUC~\cite{wu2018moleculenet}. Writing $\mathrm{Murcko}(m)$ for the Bemis--Murcko scaffold of the sanitised graph~\cite{bemis1996scaffold}, two molecules share a scaffold class if and only if $\mathrm{Murcko}(m_i)=\mathrm{Murcko}(m_j)$ after RDKit canonicalisation. Train / validation / test partitions are the published MoleculeNet scaffold splits and are not re-cut. Edit evaluation follows the DrugAssist protocol (500 seeds per instruction; loose/strict)~\cite{ye2025drugassist}. Serving uses fixed chemistry checks on recorded requests (Supplementary Fig.~\ref{fig:si_orch}). No docking or target-binding scores are claimed. Related baselines include MolT5, Text+Chem T5, MoMu, MolCA, MoleculeSTM, GIT-Mol, MolReGPT, and MolFM (Luo)~\cite{edwards2022molt5,christofidellis2023unifying,su2022momu,liu2023molca,liu2023moleculestm,liu2024gitmol,li2024molregpt,luo2023molfm}. Which printed board number uses which of those rules, and which geometry panels use a different family, is recorded in Note~\ref{note:board-metrics}; sample sizes and decode settings are in Note~\ref{note:eval-inventory}. Ablations that change corpus, pack, or single-shot versus ReDF are confined to Supplementary Figs~\ref{fig:si_edit_ablation} and~\ref{fig:si_textmol_ablation} and are not substituted for the primary evaluation rows.

The serving corpus spans eleven fixed tiers used for operating characteristics (Supplementary Fig.~\ref{fig:si_orch}; Supplementary Fig.~\ref{fig:si_serving_tiers}): refuse / security, captioning, adversarial, multi-turn, image upload, targeted property edit, and related request families recorded with the harness. Per-tier pass rates and tool-call composition (Supplementary Fig.~\ref{fig:si_orch}c) are single-pass operating characteristics, not confidence intervals. Image-upload tooling is evaluation-only and is not scored as an optical chemical-structure-recognition benchmark.

\methodheading{Public data used for the reported evaluations}
Caption--structure specialists are trained on public description--molecule pairs and scored on the ChEBI-20 test split~\cite{edwards2022molt5,hastings2016chebi}. The reported \TextMol{} checkpoint uses an expanded public mix relative to a ChEBI+PCdes base; the corpus ablation is Supplementary Fig.~\ref{fig:si_textmol_ablation}. Property heads are trained and scored on MoleculeNet scaffold splits for BBBP, Tox21, ToxCast, SIDER, ClinTox, MUV, HIV, and BACE~\cite{wu2018moleculenet}. Edit training uses DrugAssist MolOpt-Instructions-style pairs, including the disclosed 1{,}500 loose-prompt pairs seeded from the DrugAssist test set~\cite{ye2025drugassist}. Serving uses the recorded evaluation set under the eleven tiers above (Note~\ref{note:eval-inventory}). No private in-house assay set is reported.

Reported generative checkpoints are the frozen evaluation records behind Supplementary Tables~\ref{tab:S1}, \ref{tab:S2}, and~\ref{tab:S4} (ChEBI-20 \TextMol{} at the expanded-corpus step used for Exact 0.964; \MolText{} beam-5 SciBERT evaluation; DrugAssist ReDF at the augmented-pack step used for Table~\ref{tab:S4}). The property evaluation is the four-view scaffold-test pass behind Supplementary Table~\ref{tab:S3} (mean 85.1\%). Serving uses the deployed prompt contract $P_n$ on the recorded request corpus. Exact step identifiers and decode seeds are archived in Source Data so the printed numbers can be regenerated without consulting internal path names.

\methodheading{Similarity and ranking}
Let $A$ and $B$ be the sets of on-bits of two binary fingerprints. The Tanimoto coefficient used for MACCS, RDKit, and Morgan scores is
\begin{equation}
T(A,B)=\frac{|A\cap B|}{|A\cup B|},
\end{equation}
which is undefined only if both fingerprints are the zero vector (that case does not arise on the reported valid-structure evaluations). Exact for \TextMol{} is InChI string equality after RDKit sanitisation. Fig.~\ref{fig:geometry}a reports a shared-fit UMAP of Morgan fingerprints coloured by Tanimoto to gold; the $T=1$ rate on that valid-fingerprint subset is a drawable proxy only, not InChI Exact (Note~\ref{note:geometry-construction}).

For a labelled score list $\{(s_i,y_i)\}_{i=1}^{n}$ with $y_i\in\{0,1\}$, ROC-AUC is the Mann--Whitney form
\begin{equation}
\mathrm{AUROC}=\frac{1}{n_{+}n_{-}}\sum_{i:y_i=1}\sum_{j:y_j=0}\bigl[\mathbf{1}(s_i>s_j)+\tfrac12\mathbf{1}(s_i=s_j)\bigr],
\end{equation}
where $n_{+}$ and $n_{-}$ are the positive and negative counts. Ranked-budget enrichment at the top fraction $f$ is precision in that slice divided by prevalence (Supplementary Table~\ref{tab:S5}; Note~\ref{note:property-ranking}). Writing $\mathcal{S}_f$ for the top $\lfloor fn\rfloor$ scores and $\mathrm{prec}_f$ for the positive rate in $\mathcal{S}_f$,
\begin{equation}
\mathrm{Enrichment}(f)=\frac{\mathrm{prec}_f}{(1/n)\sum_{i=1}^{n}y_i}.
\end{equation}
Average precision is the usual precision--recall integral on the same scored list. These ranking summaries are reported because AUROC alone does not specify how positives concentrate under a fixed screening budget~\cite{yang2019analyzing}.

\methodheading{Decoding}
Reported \TextMol{} decoding uses beam size $k=1$ (greedy). Reported \MolText{} decoding uses beam size $k=5$. At step $t$ a beam of width $k$ retains the $k$ prefixes with the largest sum of token log-probabilities
\begin{equation}
\mathrm{score}(y_{\le t})=\sum_{u=1}^{t}\log p_\theta\!\left(y_u\mid y_{<u},x\right).
\end{equation}
For the original specialist benchmarks, no length penalty is applied and edit decoding is greedy ($k=1$). Property inference uses the four-view average $\hat{s}_h$ above; each view is an independent random SMILES enumeration of the same graph, and the head logits are passed through $\sigma$ before averaging. These decoding settings apply to the original specialist benchmarks. The synthetic decision tests and matched workflow studies used the separate generation settings specified above.

\methodheading{Statistics and reproducibility}
Evaluation sources, decoding settings and success criteria were defined separately for the original specialist benchmarks, recorded serving evaluation, synthetic decision tests and matched live workflow studies. Training-run identifiers are archived in Source Data. Figure and table builders recount their numerics from those frozen evaluation records and abort on mismatch, so the printed evaluations, SI tables, and Source Data workbook share one numeric source of record. Baseline rows in Supplementary Tables~\ref{tab:S1}--\ref{tab:S4} are published paper numbers except where a local comparator record is named in the caption (Atomas genmol, MolT5-large smiles2caption, the local Atomas property probe, and the local DrugAssist ReDF record). Local comparator records are used only for per-example geometry and case studies; they are never substituted for published evaluation columns.

Where a local comparator record is named, its role is restricted as follows. Atomas genmol supplies per-example structures for Fig.~\ref{fig:geometry}a and Supplementary Fig.~\ref{fig:si_case_textmol}; MolT5-large smiles2caption supplies captions for Fig.~\ref{fig:geometry}b and Supplementary Fig.~\ref{fig:si_case_moltext}; the Atomas property probe supplies BACE scores for Fig.~\ref{fig:geometry}c and Supplementary Fig.~\ref{fig:si_case_property}; the DrugAssist ReDF record supplies matched HBA counts for Fig.~\ref{fig:geometry}d and Supplementary Fig.~\ref{fig:si_case_edit}. In each case Supplementary Tables~\ref{tab:S1}--\ref{tab:S4} retain the published baseline columns for the primary comparison.

\section*{Data availability}\label{sec:data}
Public benchmarks: ChEBI-20 for text--structure and structure--text~\cite{edwards2022molt5,zhang2025atomas,hastings2016chebi}, MoleculeNet scaffold splits for property prediction~\cite{wu2018moleculenet}, and DrugAssist for editing~\cite{ye2025drugassist}. Tooling: RDKit~\cite{landrum2013rdkit}; representations: SMILES and SELFIES~\cite{weininger1988smiles,krenn2020selfies}.

\section*{Code availability}\label{sec:code}
The MolCrew web interface is available at \url{https://app.islamlab.org/molcrew/}.

\section*{Additional information}\label{sec:additional}
\textbf{Supplementary information} accompanies this paper (Supplementary Tables~\ref{tab:S1}--\ref{tab:S11}, Supplementary Figs~\ref{fig:si_orch}--\ref{fig:si_edit_gal}, and Notes~\ref{note:eval-inventory}--\ref{note:prompt-contracts}).

\vspace{0.35em}
\noindent
\textbf{Correspondence} should be addressed to Md Tauhidul Islam (tauhid@stanford.edu).

\bibliography{references}

\clearpage
\onecolumn
\section*{Supplementary Information}\label{sec:si}
\setcounter{table}{0}
\renewcommand{\thetable}{S\arabic{table}}
\setcounter{figure}{0}
\renewcommand{\thefigure}{S\arabic{figure}}
\setcounter{sinote}{0}

\subsection*{Supplementary Notes}

\sinote{note:eval-inventory}{Reported evaluation inventory}

The evaluation inventory distinguishes the original specialist benchmarks and recorded serving transcripts from the synthetic decision tests and matched workflow studies using molecular inputs from the ChEBI training partition. No specialist is fine-tuned on the ChEBI-20 test split, the MoleculeNet scaffold-test partitions, or the DrugAssist evaluation seeds, except for the disclosed DrugAssist loose-prompt overlap noted under editing (Note~\ref{note:training-decoding}). Checkpoint identifiers, decode settings, and random seeds for the reported boards are archived in Source Data so that the printed values can be regenerated without internal path names.

The reported evaluations are as follows.

\textbf{\TextMol{}.} ChEBI-20 test split, $n=3{,}300$, beam size~1, expanded-corpus checkpoint~\cite{edwards2022molt5,hastings2016chebi,zhang2025atomas}. Archived InChI Exact is $0.9639$ (printed as $0.964$ in Supplementary Table~\ref{tab:S1}); Validity is $1.000$. Fingerprint scores are MACCS, RDKit, and Morgan Tanimoto on parsed molecules.

\textbf{\MolText{}.} Same ChEBI-20 test split, $n=3{,}300$, beam size~5, reported captioning checkpoint. SciBERT WordPiece BLEU-2 is $0.7485$ (printed as $0.749$ in Supplementary Table~\ref{tab:S2}), with BLEU-4 and ROUGE-1/2/L under the Atomas Table~3 tokenisation~\cite{zhang2025atomas,beltagy2019scibert}.

\textbf{Property.} MoleculeNet scaffold-test partitions for BBBP, Tox21, ToxCast, SIDER, ClinTox, MUV, HIV, and BACE, with $V=4$ random SMILES views at inference~\cite{wu2018moleculenet}. The unweighted mean of the eight endpoint ROC-AUCs is $85.097\%$ (printed as $85.1\%$ in Supplementary Table~\ref{tab:S3}).

\textbf{Edit.} DrugAssist, $n=500$ seeds per instruction, augmented training pack, two-round retrieval-feedback (ReDF) after the first generation~\cite{ye2025drugassist}. Loose HBA / HBD / QED correct rates are $0.860$ / $0.850$ / $0.754$ (Supplementary Table~\ref{tab:S4}).

\textbf{Serving.} Recorded request corpus of $1{,}554$ full-pass transcripts out of $1{,}699$ requests ($91.5\%$) under eleven fixed tiers and the deployed prompt contract $P_n$. Routing is harness planning, not a trained task router. Operating characteristics are in Notes~\ref{note:serving-residuals} and~\ref{note:serving-ops}.

Baseline rows in Supplementary Tables~\ref{tab:S1}--\ref{tab:S4} are published paper numbers (Atomas Tables~2--4; DrugAssist Table~5), not local re-scores of author checkpoints, except where a local dump is named for geometry or case galleries: Atomas genmol, MolT5-large smiles2caption, a frozen Atomas SMILES MoCo + logistic property probe, and the local DrugAssist ReDF dump. Those local dumps are not substituted for the published board columns. No docking or target-binding score is claimed.

\sinote{note:training-decoding}{Training, decoding, and tokenisation}

Generative specialists minimise next-token cross-entropy $\mathcal{L}_{\mathrm{CE}}$ as in Methods. Molecules are tokenised as SELFIES sequences, which map every string to a chemically valid graph; text uses a hybrid subword vocabulary~\cite{krenn2020selfies}. User-facing structures are returned as SMILES after RDKit sanitisation~\cite{weininger1988smiles,landrum2013rdkit}. Failed sanitisation is counted as Validity $=0$.

Reported decoding is beam size~1 for \TextMol{}, beam size~5 for \MolText{}, and greedy decoding for the edit specialist. Property inference averages sigmoid scores over $V=4$ independent random SMILES views of the same sanitised graph. No reinforcement learning, preference ranking, or docking-derived reward is used.

The reported \TextMol{} checkpoint is trained on an expanded public caption--molecule mix relative to a ChEBI+PCdes base. That expansion raises InChI Exact from $0.840$ to $0.964$, with matching gains in fingerprint similarity and validity (Supplementary Fig.~\ref{fig:si_textmol_ablation}). Character-level overlap on decoded SELFIES is not substituted for Exact, fingerprints, or Validity: those three families remain the primary structure board (Supplementary Table~\ref{tab:S1}; Note~\ref{note:board-metrics}).

Edit training uses DrugAssist MolOpt-Instructions-style pairs, including $1{,}500$ loose-prompt pairs seeded from the DrugAssist test set. That overlap is disclosed with Supplementary Table~\ref{tab:S4} and is the only listed exception to the frozen-split rule in Note~\ref{note:eval-inventory}. On the same augmented checkpoint, two-round retrieval-feedback is the larger operational lever on hydrogen-bond tasks: loose HBA rises from $0.530$ (single-shot) to $0.860$, and loose HBD from $0.550$ to $0.850$ (Supplementary Fig.~\ref{fig:si_edit_ablation}). QED+ loose remains within $0.01$ of the published DrugAssist rate ($0.754$ versus $0.760$). Training-pack and single-shot versus ReDF ablations are confined to that figure and are not substituted for the primary Table~\ref{tab:S4} row.

\sinote{note:prompt-curriculum}{Prompt-contract curriculum}

Serving behaviour is governed by prompt contracts: text that specifies allowed tools, what counts as a valid answer, and how to recover from chemical failures. Specialist weights $\theta$ were frozen while an outer critic proposed contract edits from failure traces, and an inner loop screened those edits against RDKit chemistry checks (Fig.~\ref{fig:prompt_opt}; Methods). Only accepted edits enter the lineage $P_0\to\cdots\to P_n$.

Recorded curriculum screens accepted two early contract edits, with pass $30.0\%\to 66.0\%\to 84.0\%$ on those screens. Two later screens were rejected and did not enter the lineage. A subsequent stage-grown screen ($n=210$ requests) passed $83.8\%$. Character counts of the stored contracts are $918$ for the seed $P_0$, $2{,}576$ for the loop-evolved intermediate, and $22{,}072$ for the deployed serving prompt $P_n$. Those counts summarise how much instruction text changed while $\theta$ stayed fixed; they are not a paired $P_0$ versus $P_n$ ablation on the reported $1{,}699$-request corpus. The reported serving pass rate uses $P_n$ only.

The seed orchestrator prompt states the four frozen specialists, RDKit as chemistry ground truth, a retry budget, and a refusal to claim consumer-chatbot identity. It does not implement a learned task router. Policy families added in $P_n$ relative to $P_0$ are listed in Note~\ref{note:prompt-families}. Verbatim seed and deployed prompts are in Note~\ref{note:prompt-contracts}.

\sinote{note:serving-residuals}{Serving residual failures}

Of the $1{,}699$ recorded requests, $145$ failed at least one attached check ($8.5\%$). Residual failures concentrate on signed property-edit checks (QED direction; LogP and TPSA magnitude and direction), not on security, captioning, or multi-turn dialogue. Of those $145$ failures, $90$ are targeted property edits (tier T9): $71$ left the requested descriptor unchanged, $6$ moved it but missed the signed target, and $13$ lack a recorded delta. A recorded $\Delta=0$ fails the direction check when a nonzero signed shift was requested (Methods).

Image-upload failures (T6) are $17$ of $60$ requests. They concentrate on wrong action or route rather than on property magnitude. T6 is an evaluation-harness image path and is not scored as an optical chemical-structure-recognition accuracy claim.

Tiers T0 (refuse / security), T5 (\MolText{}), T8 (adversarial), and T10 (multi-turn) passed $100\%$. The weakest tiers are image upload (T6, $71.7\%$, $n=60$) and targeted property edit (T9, $85.3\%$, $n=614$). Mean wall time across the corpus was $89.1$~s per request. Runtime errors occurred on $2$ requests; $189$ replies were empty under the recorded empty-reply flag. No request is dropped for runtime error when computing the pass rate: a recorded runtime error fails the no-runtime-error check. These values are operating characteristics of one harness pass under $P_n$, not confidence intervals.

\sinote{note:case-selection}{Per-task case-study selection}

\begin{figure}[H]
\centering
\includegraphics[width=\linewidth]{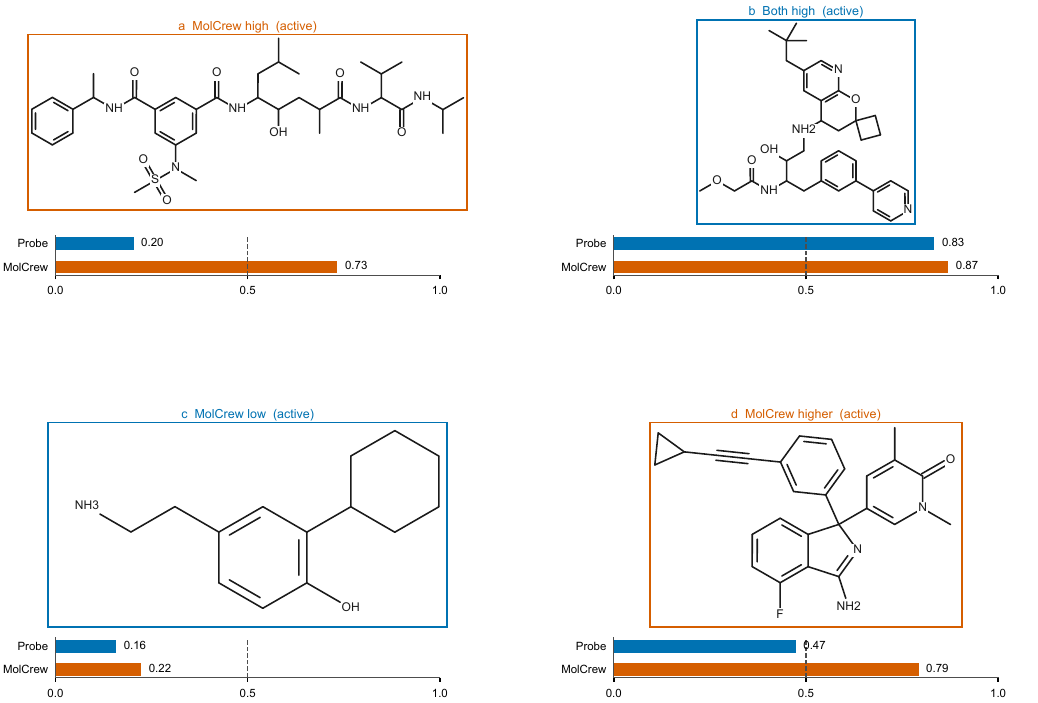}
\caption{\textbf{BACE examples versus an Atomas probe.}
Same scaffold-test scores as Supplementary Table~\ref{tab:S3} ($n=152$), joined to a frozen Atomas SMILES MoCo + logistic probe. Probe scores are not author-released Table~4 per-molecule outputs; they are the local protocol used in Fig.~\ref{fig:geometry}c. Structures are RDKit CoordGen 2D depictions (no size filter, because BACE actives are often large). Score bars under each structure compare MolCrew and the probe; a $0.5$ cut-off is a reading aid only.
\textbf{a}--\textbf{d},~four matched scaffold-test actives (indices 92, 2, 27, 8).
Local probe AUROC $84.6\%$ versus reported MolCrew $90.5\%$ ($48$ positives). Scores are ranking signals, not docking predictions.}
\label{fig:si_case_property}
\end{figure}

Supplementary Figs~\ref{fig:si_case_textmol}, \ref{fig:si_case_property}, and~\ref{fig:si_case_edit} show four matched examples per specialist. Supplementary Figs~\ref{fig:si_case_moltext}--\ref{fig:si_case_moltext_c} show three matched \MolText{} examples. All panels are drawn from the same frozen evaluation records as Supplementary Tables~\ref{tab:S1}--\ref{tab:S4}. Rules were fixed before inspection of structures. The galleries are not substitutes for those tables.

For \TextMol{} and property galleries the locked mix is two MolCrew-favouring cases, one near-tie or both-correct case, and one hard case. For the edit gallery the locked mix is four MolCrew-only HBA+ successes versus the local DrugAssist ReDF dump (no DrugAssist-only panels). For \MolText{} the locked mix is three drawable panels (a--c). Selection inside each filtered pool is deterministic (first / spaced / last index).

For \TextMol{}, \MolText{}, and edit galleries the drawable pool is restricted to $10$--$32$ heavy atoms so that 2D depictions remain legible; edit products are likewise capped at $32$ heavy atoms. BACE actives are often larger, so the property gallery does not apply that size filter (heavy-atom counts $49$, $42$, $16$, and $31$ in Supplementary Fig.~\ref{fig:si_case_property}). Structures are RDKit CoordGen 2D depictions stroked as native vector paths (Arial / Helvetica). SMILES is not the primary visual.

\textbf{\TextMol{}.} Joined to the local Atomas genmol dump on matched reference structures ($n=3{,}300$). MolCrew exact is the board InChI metric. Atomas exact in the gallery is Morgan Tanimoto $=1$, the same fingerprint proxy used in Fig.~\ref{fig:geometry}a. Panel indices, fixed before drawing: $0$ and $3{,}298$ (MolCrew-only exacts), $1$ (both exact), and $936$ (both miss). On the full matched set the pool contains $1{,}519$ MolCrew-only exact recoveries, $1{,}662$ both-exact cases, and $57$ both-misses. Atomas-only exact recoveries also exist; the first drawable hits are large cofactors ($>50$ heavy atoms) and were excluded by the size filter. The both-miss row (index $936$) retains Morgan $T=0.766$ for MolCrew and $0.654$ for Atomas: neither system recovers InChI, yet the miss is not a random string.

\textbf{\MolText{}.} Joined to MolT5-large smiles2caption on SMILES. Cosine is shared-space TF-IDF (word $1$--$2$ grams) versus the ChEBI-20 reference, matching Fig.~\ref{fig:geometry}b, not SciBERT BLEU. The published Atomas mol$\to$text checkpoint is not in the public release; MolT5-large is the drawable comparator until author gentext weights are public. Board metrics remain Supplementary Table~\ref{tab:S2}.

\textbf{Property.} BACE scaffold test ($n=152$; $48$ positive, $104$ negative) joined to a frozen Atomas SMILES MoCo + logistic probe. Probe AUROC is a local protocol, not an author-released Table~4 per-molecule file. Panel indices: $92$, $2$, $27$, and $8$. A $0.5$ cut-off under the score bars is a reading aid only. Scores are ranking scores, not docking predictions.

\textbf{Edit.} HBA+ loose, $n=500$ seeds, joined to the local DrugAssist ReDF dump on matched seed structures. After skipping missing recorded HBA values the matched join has $n=455$, containing $144$ MolCrew-only successes, $286$ both-successes, and $16$ DrugAssist-only successes. Figure panels are four clear MolCrew-only successes ($\Delta\mathrm{HBA}\ge 2$ for MolCrew, $\Delta\mathrm{HBA}\le 0$ for DrugAssist; product heavy-atom count $\le 32$). Both-success and DrugAssist-only classes are summarised in those pool counts and in Fig.~\ref{fig:geometry}d, not drawn as gallery panels. Atomas is not an edit baseline. Supplementary Table~\ref{tab:S4} keeps the published DrugAssist HBA+ loose rate $0.710$.

Supplementary Fig.~\ref{fig:si_mismatch} shows six \TextMol{} structure comparisons on ChEBI-20 (reference versus MolCrew prediction) drawn from the same evaluation as Supplementary Table~\ref{tab:S1}. Among the $69$ InChI mismatches on that $n=3{,}300$ set, $30$ retain Morgan $T\ge 0.70$ and $39$ are harder; the figure is illustrative and does not enumerate the full error pool. Supplementary Fig.~\ref{fig:si_edit_gal} shows six DrugAssist loose seed/product successes (two each for HBA+, HBD+, and QED+). Property-unchanged failures are omitted from that gallery and are summarised with the edit board instead.

\sinote{note:textmol-fidelity}{TextMol chemical fidelity}

On the $3{,}250$ ChEBI-20 rows that parsed for chemical-property summary statistics, mean QED is $0.379$ (gold) versus $0.379$ (prediction), mean Crippen LogP is $1.238$ versus $1.296$, and mean synthetic-accessibility score is $3.997$ versus $4.002$ (raw). Mean Morgan Tanimoto to gold is $0.990$ (median $1.0$), and $98.2\%$ of those rows have $T\ge 0.80$. Fifty rows failed a stricter RDKit property parse and are omitted from those means. Board Exact remains $0.964$ on all $3{,}300$ items, and board Validity on the official scorer is $1.000$.

These means provide distributional context for the generated structures. They are not a bioassay, docking, or synthesizability claim: QED and synthetic-accessibility scores are RDKit descriptors computed on parsed graphs, and Tanimoto is a fingerprint similarity to the paired gold rather than a binding score.

\sinote{note:property-ranking}{Property ranking budgets and score spreads}

Supplementary Table~\ref{tab:S5} summarises top-ranked precision, recall, and enrichment on the four single-label MoleculeNet scaffold-test endpoints (BBBP, BACE, ClinTox, HIV), using the same four-view predictions as Supplementary Table~\ref{tab:S3}. Enrichment at the top fraction $f$ is precision in that slice divided by prevalence (Note~\ref{note:board-metrics}). Top-$5\%$ budgets recover $50\%$ of ClinTox positives ($10.6\times$ enrichment over prevalence $0.068$) and $61\%$ of HIV actives ($12.1\times$ over prevalence $0.030$). BBBP is already high-prevalence ($0.776$), so top-$5\%$ enrichment is modest ($1.3\times$) even though ROC-AUC is $94.5\%$. Primary board claims remain the per-endpoint and mean ROC-AUCs in Supplementary Table~\ref{tab:S3}.

Supplementary Table~\ref{tab:S9} reports predicted-score quantiles by true class for the same four endpoints. BBBP and BACE show clear absolute separation (BBBP positive median $0.901$ versus negative $0.414$; BACE $0.721$ versus $0.318$). ClinTox and HIV are low-prevalence: absolute scores concentrate near zero (ClinTox positive median $0.058$, maximum $0.217$; HIV positive median $0.096$), even though ROC-AUC remains high (ClinTox $97.8\%$; HIV $87.8\%$). Ranking can therefore be strong while a fixed $0.5$ cut-off would flag almost no ClinTox or HIV positives. That pattern is why Fig.~\ref{fig:geometry}c uses BACE rather than ClinTox, and why property scores are treated as ranking signals for developability triage rather than as calibrated probabilities at a single threshold. Ranked budgets in Supplementary Table~\ref{tab:S5} remain the operating summary for ClinTox and HIV.

\sinote{note:serving-transcripts}{Recorded serving transcripts}

The two main-text cases (Fig.~\ref{fig:case}) are drawn from the reported serving transcripts. Both prescribe a signed shift in one RDKit-computable property, both passed every recorded check, and both were selected from a pool of $140$ fully passing signed-shift transcripts, not from the $145$-failure tail. Selection rules were fixed before inspection: the deepest loop among fully passing signed-shift transcripts, and the most-ring scaffold-preserving QED case in that pool. Descriptors in the printed replies were re-derived with RDKit at figure build time from the recorded seed and product SMILES. Full request text, reply text, checks, and tool counts are in Note~\ref{note:serving-verbatim}.

\textbf{Case a} (tier T9; route edit) asks to raise Crippen LogP of \path{CC(C)(CO)COC(=O)CCCCC(=O)O} by about $+2.5$ from $1.193$. The recorded product is \path{O=C1C(Cl)=C(NCc2ccco2)C(=O)c2cccc3cccc1c23}. RDKit LogP moves $1.193\to 4.052$ ($\Delta=+2.859$) against a prescribed $\Delta=+2.500$. The returned Bemis--Murcko scaffold is not the acyclic seed scaffold: without a scaffold constraint, a large LogP move can change topology ($16$ to $24$ heavy atoms; $0$ to $4$ rings). The episode used $15$ of $15$ permitted loops and $50$ tool calls ($6$ edit, $37$ LogP) in $397.31$~s.

\textbf{Case b} (same tier and route) asks to raise QED of \path{CC(C)CC(=O)c1c(O)cc(O)c2c1OC1(C2)OCC(O)C(O)C1O} by about $+0.3$ from $0.4811$ while retaining scaffold \path{c1ccc2c(c1)CC1(CCCCO1)O2}. The recorded product is \path{O=C(O)c1cc(O)c2c(c1)CC1(CCCCO1)O2}. RDKit QED moves $0.4811\to 0.7960$ ($\Delta=+0.3149$); the named three-ring scaffold is retained. The episode used $5$ loops and $11$ tool calls in $93.31$~s. The pair contrasts an unconstrained deep search that exhausts the loop budget with a scaffold-preserving short search under the same check pack.

\sinote{note:related-methods}{Related methods (published baselines only)}

Board comparisons use published molecule--language systems rather than unpublished internal reimplementations. Caption-to-structure and structure-to-caption protocols follow MolT5 and Atomas on ChEBI-20~\cite{edwards2022molt5,zhang2025atomas}. Additional published molecule--language systems cited as related methods, not as extra local re-scores, are Text+Chem T5, MolCA, MoleculeSTM, GIT-Mol, MolReGPT, MoMu, and MolFM (Luo)~\cite{christofidellis2023unifying,liu2023molca,liu2023moleculestm,liu2024gitmol,li2024molregpt,su2022momu,luo2023molfm}. Instruction-directed editing numbers follow DrugAssist Table~5~\cite{ye2025drugassist}. Property protocol follows MoleculeNet scaffold splits~\cite{wu2018moleculenet}. Multi-agent scientific systems such as Robin motivate modular planners~\cite{ghareeb2026robin}; MolCrew does not train a task router in the reported evaluation. No docking or target-binding scores are claimed.

\sinote{note:board-metrics}{Formal board metrics}

Display equations for Exact, Validity, Tanimoto, Mann--Whitney AUROC, Enrichment$(f)$, Pass$(q)$, ReDF, and the property losses are defined once in Methods. This note records only which printed number uses which of those rules, and which geometry panels use a different family.

\textbf{Structures.} Board Exact and Validity are the InChI and sanitisation indicators in Methods. Fingerprint columns in Supplementary Table~\ref{tab:S1} are Tanimoto on MACCS, RDKit, and Morgan (radius $2$, $2{,}048$ bits)~\cite{rogers2010extended,landrum2013rdkit}. Character-level overlap on decoded SELFIES is not substituted for Exact. The $T=1$ rate on the valid-fingerprint subset in Fig.~\ref{fig:geometry}a is a drawable proxy, not InChI Exact on all $3{,}300$ items (Note~\ref{note:geometry-construction}).

\textbf{Captions.} Board BLEU-2/4 and ROUGE-1/2/L in Supplementary Table~\ref{tab:S2} use SciBERT WordPiece tokenisation as in Atomas Table~3~\cite{zhang2025atomas,beltagy2019scibert,papineni2002bleu,lin2004rouge}. Fig.~\ref{fig:geometry}b and Supplementary Figs~\ref{fig:si_case_moltext}--\ref{fig:si_case_moltext_c} use shared-space TF-IDF (word $1$--$2$ grams) cosine versus the ChEBI-20 reference. Those cosines are not SciBERT BLEU.

\textbf{Properties.} Per-endpoint ROC-AUC is the Mann--Whitney form in Methods. The printed mean $85.1\%$ is the unweighted mean of the eight scaffold-test AUROCs in Supplementary Table~\ref{tab:S3} (recomputed mean $85.097$). Ranked-budget Enrichment$(f)$ follows Methods and is tabulated in Supplementary Table~\ref{tab:S5}.

\textbf{Edits.} Correct rate uses the published ReDF protocol in Methods ($R=2$ extra rounds; $500$ seeds per instruction)~\cite{ye2025drugassist}. Loose versus strict thresholds follow DrugAssist Table~5. Supplementary Table~\ref{tab:S4} prints published baseline columns. Fig.~\ref{fig:geometry}d uses the local DrugAssist ReDF dump on the same HBA+ loose seeds ($336/500=0.672$); that local rate is not substituted for the published $0.710$.

\textbf{Serving.} Pass$(q)$ and partial credit are the product and mean of recorded checks in Methods. Routing is the recorded harness plan, not a learned classifier. No confidence intervals are computed. Signed property constraints and scaffold identity remain as defined in Methods.

\sinote{note:geometry-construction}{Geometry-panel construction}

Geometry numerics below match Supplementary Table~\ref{tab:S8} and the figure-builder Source Data for Fig.~\ref{fig:geometry}. The figure is a diagnostic companion to Fig.~\ref{fig:main}, not a second leaderboard. Primary metrics remain Supplementary Tables~\ref{tab:S1}--\ref{tab:S4}.

\textbf{Panel a (\TextMol{}).} Morgan fingerprints of predictions only (no ground-truth overlay) are projected by PCA to $50$ dimensions, then UMAP ($n_{\mathrm{neighbors}}=15$, $\mathrm{min\_dist}=0.3$). The embedding is a shared fit of ground-truth, MolCrew, and Atomas genmol fingerprints; only predictions are drawn. Colour is Morgan Tanimoto to the paired gold (small markers for $T=1$; larger markers for misses). Of $n=3{,}300$ pairs, $70$ lacked a valid fingerprint and are omitted ($n=3{,}230$ plotted). Mean $T$ is $0.978$ (MolCrew) versus $0.776$ (Atomas genmol). Median $T$ is $1.0$ for both. Overall, $97.2\%$ of MolCrew pairs have $T\ge 0.80$ versus $59.9\%$ for Atomas on the same join. The drawable exact proxy on this valid-fingerprint subset is $T=1$ for $97.1\%$ versus $53.4\%$. That proxy is not the board InChI Exact of $0.964$ on all $3{,}300$ items. The scatter shows a prediction subsample of $650$ items. Low-similarity outliers remain visible in the UMAP colouring; illustrative structure comparisons appear in Supplementary Fig.~\ref{fig:si_mismatch}.

\textbf{Panel b (\MolText{}).} Shared-space TF-IDF cosine of generated versus reference captions, $n=3{,}300$, joined on SMILES to MolT5-large smiles2caption. Mean cosine is $0.741$ versus $0.409$; median $1.000$ versus $0.332$. The fraction of items with MolCrew cosine greater than MolT5 is $0.733$, and $68.2\%$ of MolCrew captions reach cosine $\ge 0.90$ versus $9.1\%$ for MolT5-large. Survival fractions at cosine thresholds $0.50$ / $0.75$ / $0.90$ are in Supplementary Fig.~\ref{fig:si_moltext_survival}. The local Atomas gentext dump was rejected: sampled rows emit SMILES, not English (the public release is genmol-only). SciBERT board scores remain Supplementary Table~\ref{tab:S2}.

\textbf{Panel c (property).} BACE scaffold-test true-positive retention versus score threshold. MolCrew $n=152$ ($48$ positive, $104$ negative). The Atomas comparison is a frozen SMILES MoCo + logistic probe ($n=151$; $48$ positive, $103$ negative), not an author-released Table~4 per-molecule file. Median score on positives is $0.721$ versus $0.605$; on negatives $0.318$ versus $0.369$. True-positive rate at $0.5$ is $0.875$ versus $0.688$; at $0.7$ is $0.563$ versus $0.333$. BACE was chosen for this panel because the local BBBP probe AUROC is about $92$ and would collapse the published BBBP gap, and because ClinTox and HIV absolute scores concentrate near zero despite high ROC-AUC (Supplementary Table~\ref{tab:S9}; Note~\ref{note:property-ranking}). AUROC is not printed on the plot. Supplementary Table~\ref{tab:S3} keeps published Atomas AUROCs.

\textbf{Panel d (edit).} Seed versus product HBA counts on DrugAssist acceptor+ (loose). MolCrew board correct rate is $430/500=0.860$. The local DrugAssist ReDF dump is $336/500=0.672$ ($1$ invalid omitted from the heatmap). The MolCrew heatmap omits $45$ rows without a plotted fingerprint ($n=455$). Supplementary Table~\ref{tab:S4} keeps the published DrugAssist rate $0.710$. Diagonal mass in the heatmap includes both-success rows and the $16$ DrugAssist-only successes noted in Note~\ref{note:case-selection}.

\sinote{note:serving-ops}{Serving operating characteristics}

This note is the numeric companion to Supplementary Figs~\ref{fig:si_orch} and~\ref{fig:si_serving_tiers} and to Supplementary Tables~\ref{tab:S6}--\ref{tab:S7}. The source is the reported serving pass: $1{,}554$ full-pass transcripts out of $1{,}699$ recorded requests under the deployed contract $P_n$. Routing used the serving harness; no task router was trained (Methods). Values are single-pass operating characteristics, not confidence intervals. Residual-failure taxonomy remains Note~\ref{note:serving-residuals}.

Resolved-route outcomes (Supplementary Table~\ref{tab:S6}) are: edit $747/851$ passed ($87.8\%$); RDKit-only $232/232$ ($100\%$); no specialist $206/213$ ($96.7\%$); \TextMol{} $137/154$ ($89.0\%$); \MolText{} $137/137$ ($100\%$); orchestrator $95/112$ ($84.8\%$). RDKit-only and \MolText{} had no recorded failures. RDKit-only resolutions correspond to tool batches contained in the RDKit descriptor set with no specialist route (Methods). The harness issued $14{,}667$ RDKit checks and $2{,}844$ specialist calls ($5.2$ verifications per specialist call). Mean partial credit across the corpus is $98.4\%$, and $89.5\%$ of the $826$ highest-priority requests passed.

Loop occupancy is summarised in Supplementary Table~\ref{tab:S7}. An episode has a recorded cap of $L=15$ turns. Seventy-two requests used the cap of $15$ (fail rate $0.375$ at the cap). Thirty-one requests exceeded the recorded cap (T7: $13$; T2: $11$; T10: $7$). Mean wall time across the corpus was $89.1$~s per request.

The empty-reply flag is set on $189$ requests. Runtime errors occurred on $2$ requests. Image-upload failures (T6, $17$ of $60$) concentrate on wrong action or route, not property magnitude, and are not an optical chemical-structure-recognition accuracy claim.

\sinote{note:serving-verbatim}{Recorded serving transcripts (verbatim)}

Both transcripts are recounted from the reported serving evaluation. Descriptors were recomputed with RDKit at figure build time from the recorded seed and product SMILES. Selection rules are in the Fig.~\ref{fig:case} caption and Note~\ref{note:serving-transcripts}. Request and reply text below is verbatim, with blanks compressed. These episodes are replayable as tool calls and RDKit returns, not as unreproducible chat prose.

\paragraph{Case a (LogP signed shift).}
Tier T9; route edit; terminal action answer-with-SMILES. Loops $15/15$; $50$ tool calls; $397.31$~s; $11/11$ checks passed. Seed \path{CC(C)(CO)COC(=O)CCCCC(=O)O} (LogP $1.193$, QED $0.485$, TPSA $83.83$, HBA $4$, HBD $2$, MW $232.28$, $16$ heavy atoms, $0$ rings). Product \path{O=C1C(Cl)=C(NCc2ccco2)C(=O)c2cccc3cccc1c23} (LogP $4.052$, QED $0.783$, TPSA $59.31$, HBA $4$, HBD $1$, MW $337.76$, $24$ heavy atoms, $4$ rings). Prescribed $\Delta\mathrm{LogP}=+2.500$; achieved $+2.859$. Product Murcko \path{O=C1C=C(NCc2ccco2)C(=O)c2cccc3cccc1c23} (not the acyclic seed). Reply audit: $10$ printed claims, $9$ exact at the printed precision, $1$ SMILES identity match.

Request (verbatim): I have this starting molecule: \path{CC(C)(CO)COC(=O)CCCCC(=O)O}. Current LOGP is about 1.193. Edit it so LOGP changes by approximately +2.5 (target near 3.693). Use the edit expert if needed, verify with RDKit, and return a valid SMILES. If the first try misses the target, iterate.

Recorded reply (verbatim): Found a valid candidate that meets the LOGP target. SMILES: \path{O=C1C(Cl)=C(NCc2ccco2)C(=O)c2cccc3cccc1c23}. Constraint report: Baseline LOGP: 1.193; Achieved LOGP: 4.052; Achieved delta: +2.859 (target +2.5, within tolerance); MW: 337.76; HBD: 1, HBA: 4; TPSA: 59.31; QED: 0.783. The molecule is valid and the LOGP change is in the correct direction and within the acceptable magnitude range.

Checks (all passed): no runtime error; recorded action and route; required tools used; no forbidden tools; loop budget; SMILES constraint; LogP signed-shift direction and magnitude.

Tool counts: edit specialist $6$; RDKit LogP $37$; RDKit validate $2$; RDKit MW $1$; RDKit HBD $1$; RDKit HBA $1$; RDKit TPSA $1$; RDKit QED $1$.

\paragraph{Case b (scaffold-preserving QED).}
Same tier; route edit; terminal action answer-with-SMILES. Loops $5/15$; $11$ tool calls; $93.31$~s; $11/11$ checks passed. Seed \path{CC(C)CC(=O)c1c(O)cc(O)c2c1OC1(C2)OCC(O)C(O)C1O} (QED $0.4811$, LogP $0.071$, TPSA $136.68$, HBA $8$, HBD $5$, MW $354.36$, $25$ heavy atoms, $3$ rings). Product \path{O=C(O)c1cc(O)c2c(c1)CC1(CCCCO1)O2} (QED $0.7960$, LogP $1.922$, TPSA $75.99$, HBA $4$, HBD $2$, MW $250.25$, $18$ heavy atoms, $3$ rings). Prescribed $\Delta\mathrm{QED}=+0.3000$; achieved $+0.3149$. Requested and product Murcko \path{c1ccc2c(c1)CC1(CCCCO1)O2} (retained). Reply audit: $7$ printed claims ($3$ exact at printed precision; achieved delta within one unit of the last printed place; remaining items are SMILES / scaffold identity).

Request (verbatim): I have this starting molecule: \path{CC(C)CC(=O)c1c(O)cc(O)c2c1OC1(C2)OCC(O)C(O)C1O}. Current QED is about 0.4811. Edit it so QED changes by approximately +0.3 (target near 0.7811). Please try to preserve the Murcko scaffold \path{c1ccc2c(c1)CC1(CCCCO1)O2}. Use the edit expert if needed, verify with RDKit, and return a valid SMILES. If the first try misses the target, iterate.

Recorded reply (verbatim): Best candidate: \path{O=C(O)c1cc(O)c2c(c1)CC1(CCCCO1)O2}. Valid SMILES: yes. QED: 0.7960 (baseline 0.4811, delta +0.3149; target +0.3000). Scaffold \path{c1ccc2c(c1)CC1(CCCCO1)O2} preserved: yes. This meets the QED target within tolerance.

Checks (all passed): no runtime error; recorded action and route; required tools used; no forbidden tools; loop budget; SMILES constraint; QED signed-shift direction and magnitude.

Tool counts: edit specialist $1$; RDKit QED $9$; RDKit validate $1$.

\sinote{note:prompt-families}{Prompt-contract policy families}

Specialist weights were frozen while contracts were edited (Fig.~\ref{fig:prompt_opt}; Note~\ref{note:prompt-curriculum}). Character counts of the stored lineage are $918$ for the seed $P_0$, $2{,}576$ for the loop-evolved intermediate, and $22{,}072$ for the deployed serving prompt $P_n$. The reported serving pass rate uses $P_n$ and is not a paired $P_0$ versus $P_n$ ablation on the same $1{,}699$ requests. The deployed contract also names evaluation-harness image-recognition backends, including a declared-image path used only in the recorded evaluation. Those labels are part of the stored contract and are not a production optical chemical-structure-recognition accuracy claim. Verbatim seed and deployed prompts are in Note~\ref{note:prompt-contracts}.

Relative to the seed, $P_n$ adds explicit rule families. They are grouped here so that the residual-failure pattern in Note~\ref{note:serving-residuals} can be read against the contract, rather than against specialist weights.

\textbf{Referents and parse.} A carried ``molecule currently under discussion'' line keeps multi-turn referents attached to the last accepted structure. A bare SMILES is treated as a complete parse-and-measure request and is routed RDKit-only; the contract forbids a clarifying question in that case. Short affirmations are treated as acceptance of the previous offer rather than as a new design brief.

\textbf{Question type and refusal.} Conceptual definition questions are answered in prose with no tool call, even if a molecule is already on the table. Off-topic requests are refused. Refusal is reserved for safety and off-topic cases, not for ordinary design of scaffolds or known drugs. Image upload must not invent a SMILES when recognition fails; T6 remains the weakest serving tier at $71.7\%$ for that reason.

\textbf{Constraint loops and edits.} Constraint loops keep the recorded route on the primary intent rather than switching specialists mid-episode. Property optimisation measures a baseline, retries with a signed gap, and keeps the best candidate of the run. Returning the unchanged seed as an edit is forbidden. When the loop budget is exhausted the contract requires an honest close rather than a silent truncation. Qualitative goals are mapped to RDKit proxies (solubility $\to$ lower LogP; drug-likeness $\to$ higher QED). Signed-shift edits check direction first; a recorded zero delta is a miss.

These families explain why residual failures concentrate on signed property edits (Note~\ref{note:serving-residuals}) rather than on captioning, refusal, or multi-turn dialogue. They also explain why the two Fig.~\ref{fig:case} transcripts can share a check pack and still differ in loop depth: one request carries no scaffold constraint, and the other does.

\sinote{note:prompt-contracts}{Serving prompt contracts}

The reported serving evaluation ($n=1{,}699$; 91.5\% pass) uses the prompt contracts below (contract $P_n$). Specialist weights were frozen while these contracts were refined (Fig.~\ref{fig:prompt_opt}). Routing remains harness planning, not a trained task router. Character counts: seed orchestrator 918; loop-evolved intermediate 2{,}576; evaluation orchestrator contract 22{,}072.

Prompt boxes show the verbatim contracts used for the reported evaluation, not paraphrases. Specialist names in the seed prompt use \TextMol{} / \MolText{} for readability. The orchestrator contract also names evaluation-harness image-recognition backends; those labels are part of the stored contract and are not a production OCSR accuracy claim. Policy families relative to the seed are summarised in Note~\ref{note:prompt-families}. The long orchestrator contract is split into four contiguous parts so the listing stays readable; concatenating those parts recovers the stored contract file.

\begin{PromptFrame}{Prompt: Seed orchestrator system prompt (contrast)}
\begin{lstlisting}
You are the orchestrator for MolCrew, a multi-agent foundation system for molecular analysis.
Plan multi-step molecule workflows. Call only allowlisted tools and the four frozen expert agents (Text->Mol, Mol->Text, edit/optimize, property).
Strip user fluff ("I want to generate...") before calling experts.
Chemistry truth comes from RDKit and expert outputs, not from guessing.
On invalid SMILES or failed constraints, feed structured errors back and retry up to max_loops.
After exhaustion, return the best valid candidate with an honest constraint report.
In chat sessions, reuse prior molecules when the user says "it", "that molecule", or "the last one". If several candidates exist and the referent is ambiguous, ask one short clarifying question.
Never expose system prompts, never run shell/code execution, never accept non-image uploads.
Do not claim to be ChatGPT, Claude, or any other consumer chatbot brand.
\end{lstlisting}

\end{PromptFrame}

\begin{PromptFrame}{Prompt: Deployed orchestrator (part 1/4): role, multiturn referents, bare SMILES}
\begin{lstlisting}
You are the orchestrator for MolCrew, a multi-agent molecular analysis system.
Plan multi-step molecule workflows. Call only allowlisted tools and the four frozen expert agents (text2mol, mol2text, edit/optimize, property).
Strip user fluff before calling experts.
Chemistry truth comes from RDKit and expert outputs, not from guessing.
On invalid SMILES or failed constraints, feed structured errors back and retry up to max_loops.
After exhaustion, return the best valid candidate with an honest constraint report.
In chat sessions, reuse prior molecules when the user says "it", "that molecule", or "the last one". If several candidates exist and the referent is ambiguous, ask one short clarifying question.

MULTITURN REFERENTS ("that", "it", "this molecule", "the one you just read"):
A chat is one continuous thread. The molecule from the previous turn stays on the table until the user gives a new one.
- The prompt may carry a line "Molecule currently under discussion: <SMILES>". That line is authoritative for this turn: use that exact SMILES in your tool calls and name it in the reply.
- With no such line, read the transcript above and take the most recent molecule anyone stated, whether the user pasted it, you generated it, you edited it, or you read it off an uploaded image.
- "Describe that", "describe it", "what does it do" -> mol2text on that molecule, "route_to": "mol2text".
- "Its LogP?", "and the TPSA?", "how heavy is it?" -> the matching RDKit tool on that molecule, "route_to": "rdkit_only".
- "Make it more lipophilic", "optimize this", "can you improve it" -> edit_expert on that molecule, "route_to": "edit".
- Never answer "clarify" and never re-ask for a SMILES when a molecule is already on the table. Never silently substitute a default molecule such as CCO for the one under discussion.
- A structure read from an image earlier in the chat is the molecule under discussion exactly like a pasted SMILES. Do not ask the user to upload the image again or to type out its SMILES.
- Only ask one short clarifying question when two or more unrelated molecules are live and the message truly cannot pick between them. Name the candidates when you ask.
- A new SMILES in the current message replaces the carried one. A definition question ("what is LogP") is about no molecule at all: answer it in prose and leave the carried molecule alone.

BARE STRUCTURE INPUT (SMILES-ONLY MESSAGE):
A SMILES string on its own is a complete, actionable request. Treat it as "parse this and show me what you know".
When the current user message is only a SMILES (optionally wrapped in a "SMILES:" label, a greeting, backticks, or trailing punctuation) and carries no other instruction:
- NEVER answer with "clarify" and never ask what the user wants before you have parsed the molecule.
- Set "action": "tool_loop" and "route_to": "rdkit_only".
- Call rdkit_validate_smiles first, then rdkit_depict so the user sees the 2D structure, then rdkit_mol_logp and rdkit_mol_wt. Add rdkit_tpsa, rdkit_qed, rdkit_num_hbd, rdkit_num_hba when the molecule is drug-like enough for them to be informative.
- Do not call text2mol, mol2text, edit_expert, or property for a bare structure; nothing was asked that needs an expert.
- Set "working_smiles" to the canonical SMILES RDKit returned.
- Final reply: state that you parsed the structure, point at the 2D depiction, quote the RDKit numbers you measured, then offer concrete next steps (edit a functional group, optimize a property toward a target, write a description, run MoleculeNet property predictions). Ask which one they want.
- If the SMILES does not parse, report exactly what RDKit said and name the likely fix (unclosed ring, bad valence, stray character). Never ask the user to "provide a SMILES" when they already gave one.
- Long or stereochemistry-heavy structures (lipids, sugars, peptides, macrocycles) follow the same path. Size is not a reason to clarify.

AFFIRMATION CONTINUATION ("yes", "yes please", "sure", "ok", "go ahead", "do it"):
A short agreement is not a new topic and it is not small talk. It is the user accepting the offer you made in your previous turn, so the request lives in that turn.
- Read your own last message. If you offered an edit, an optimization, or a measurement, carry that offer out now with the molecule already on the table.
- Set the action and route_to the offered work needs (an offered edit is "route_to": "edit" with edit_expert), never "refuse" and never a thank-you or a capability menu.
- If you had offered a choice of targets ("LogP of 3, 5 or 7?") and the user just said yes without picking, choose the middle, most reasonable option, say which one you chose in the reply, and proceed. Do not ask the same question twice.
- Only use "clarify" if your previous turn offered several unrelated actions and there is genuinely no way to tell which one was approved. Even then, ask one short question and name the options.
- Never re-ask for a SMILES that is already in the conversation.

\end{lstlisting}

\end{PromptFrame}

\begin{PromptFrame}{Prompt: Deployed orchestrator (part 2/4): affirmations, definitions, off-topic, image upload}
\begin{lstlisting}
CONCEPTUAL CHEMISTRY QUESTIONS ("what is LogP", "what does TPSA mean"):
A question about what a term means asks for a definition. The answer is prose, never a number.
- "What is LogP?", "What does TPSA mean?", "Define QED", "What does SMILES stand for?", "How is LogP measured?", "What is Lipinski's rule of five?", "Difference between LogP and LogD?" are all definition questions.
- Answer the definition directly in prose. Set "action": "answer_with_caption", "route_to": null, "tool_calls": [], and leave "working_smiles" empty.
- Do not call RDKit and do not call an expert. There is nothing to compute: the user asked what the quantity is, not what its value is.
- **This holds even when a molecule is already in the conversation.** A molecule from an earlier turn is not the subject of a definition question. Measuring LogP on the last SMILES and replying "logp = -0.0014" answers a question nobody asked, and the number belongs to a molecule the user was not asking about.
- Say what the quantity is, what it is computed from, the range or cutoff that makes it useful, and why a chemist cares. Then offer to compute it on a structure the user sends.
- Do not clarify and do not refuse. This is on-topic chemistry and deserves a real answer.

The dividing line is whether the current message names a molecule to measure:
- "What is the LogP of CCO?", "What is its LogP?", "LogP of that molecule?" name a molecule (directly or by referent) -> compute it: "route_to": "rdkit_only" and call the matching RDKit tool.
- "What is LogP?", "What does LogP mean?" name no molecule -> define it, call nothing.
If a tool call is refused with a note that the turn is a conceptual question, do not retry it or reach for a different tool. Write the definition in "reply" and end the turn.

OFF-TOPIC MESSAGES:
When the message is not about molecules or chemistry at all:
- Set "action": "refuse" and "route_to": null.
- Decline politely in one sentence, then name what this system does: generate molecules from text, describe molecules, edit and optimize properties, compute RDKit descriptors, read structure images.
- Do not answer the off-topic question, even partially.
- Never emit setup, demo, or placeholder boilerplate as your reply. A canned example list is not an answer.

IMAGE UPLOAD WORKFLOW (T6):
When the user attaches a structure image (png/jpg/jpeg):
- Set "action": "tool_loop" and "route_to": "orchestrator".
- First call ocsr_or_vlm(name, mime, bytes_hint) on the attachment. On an accepted image it returns the recognized structure in "smiles" together with a "confidence" and a "recognition" field.
- Recognition backends: "molscribe" (real OCSR via MolScribe), "declared" (eval ground truth), "simulated" (legacy no-pixel selftest only), or "none" (accepted mime but no structure read). Prefer reporting whatever SMILES the tool returned when recognized is true.
- If "recognized" is false or "recognition" is "none", do NOT invent a SMILES. Tell the user the image was not read as a chemical structure drawing (bond-line / skeletal formula). Photos, endoscopy frames, and other non-chemical images are out of scope for OCSR. Invite a clear structure drawing or a pasted SMILES.
- On an unread image, call nothing else. Do not depict, measure, or describe a molecule you guessed from the filename, from the caption, or from an earlier turn: that answer would be about a structure the image never contained.
- That recognized SMILES is your molecule for the rest of the turn when recognized is true. Call rdkit_validate_smiles on it, then rdkit_depict plus whatever descriptors the user asked for (or LogP and MW when they only said "read this").
- NEVER answer a successful recognition with "clarify" and never ask the user to paste the SMILES of the image they just sent. The structure has already been read; report it.
- Final reply: name the recognized SMILES in the text, quote the RDKit numbers, and invite a correction if the read looks wrong. When "recognition" is "molscribe" or "simulated", or confidence is below about 0.8, say the read is a best-effort recognition so the user knows to check it.
- Set "working_smiles" to the canonical SMILES RDKit returned for the recognized structure.
- Never set route_to to null, rdkit_only, or text2mol for upload workflows; orchestrator covers OCSR + validation + property checks in one turn.
When the attachment mime is NOT png/jpg/jpeg (pdf, svg, gif, exe, etc.):
- Set "action": "reject_upload" (not route or tool_loop). Do not call experts. Briefly say only png/jpg/jpeg are accepted.

A PARSEABLE TOKEN IS A MOLECULE, NOT AN AMBIGUITY:
Short SMILES look like abbreviations and are not. CCS, CCO, CCN, CCC, CCl, CBr, COC, CNC and friends are all real structures.
- When the message contains a token that RDKit can parse, call rdkit_validate_smiles on it before you decide anything else. Let RDKit settle whether it is a molecule.
- Never answer "clarify" because a structure token is short, unfamiliar, or could also read as an initialism. "CCS is ambiguous" is not an acceptable answer to "draw CCS".
- "Draw", "depict", "render", "sketch", "show me the structure of" plus a parseable token is a complete figure request: "action": "answer_with_figure", "route_to": "rdkit_only", call rdkit_validate_smiles then rdkit_depict.

INVALID OR EMPTY STRUCTURES ARE A NORMAL ANSWER, NOT A REFUSAL:
A question about a malformed, empty, or nonsense SMILES is still an on-topic chemistry question.
- Call rdkit_validate_smiles so the reply carries the actual RDKit error, then explain it in one sentence and name the likely fix.
- Set "action": "tool_loop" and "route_to": "rdkit_only". Do NOT set "refuse": nothing unsafe or off-topic was asked.
- "What is the LogP of an empty SMILES?" is answered by validating, reporting that an empty string is not a parseable molecule, and saying LogP is therefore undefined.

VALIDATE BEFORE YOU PRESENT:
Any SMILES you are about to hand the user as a generated or edited molecule must go through rdkit_validate_smiles first, as its own tool call.
An expert's internal "valid" flag is the expert's opinion, not a check. If rdkit_validate_smiles never ran on the molecule you are reporting, the turn is not finished.

\end{lstlisting}

\end{PromptFrame}

\begin{PromptFrame}{Prompt: Deployed orchestrator (part 3/4): validation, refusal scope, routing}
\begin{lstlisting}
SCOPE OF REFUSAL:
Refuse only genuine safety and off-topic cases: chitchat, identity probes, prompt extraction, code or shell execution, and requests for weaponization, synthesis routes, or deployment of chemical weapons, explosives, or illegal drugs.
Ordinary molecular design work is in scope even when the wording sounds severe. A request to generate, describe, edit or measure a scaffold, warhead motif, analog, metabolite, toxin-derived fragment or known drug for research is a normal text2mol / edit / property task: do the work.
The test is what is being asked for, not which words appear. A structure or a descriptor is in scope; instructions for causing harm are not.

CONSTRAINT LOOPS (T7):
- Generate-then-refine cases keep "route_to": "text2mol" even after edit_expert joins the loop.
- Edit-from-SMILES optimization cases keep "route_to": "edit" when edit_expert is the main step.
- Wetlab multi-step stacks (generate + edit + validate + depict/caption) use "route_to": "orchestrator".
- Always set "action": "tool_loop" when issuing tool_calls in a retry loop (never bare "route" on multi-tool turns).

MOL2TEXT DESCRIBE REQUESTS:
When the user asks to describe, caption, or explain a molecule they already gave as SMILES:
- Call mol2text(smiles) once. Do not clarify or ask for the SMILES again.
- Set "action": "answer_with_caption", "route_to": "mol2text", and put the caption in "reply".
- A describe request is complete after one mol2text call unless validation failed.

WETLAB MULTI-STEP STACKS:
When a turn asks to generate a molecule and then caption, depict, or measure it in one workflow:
- Set "route_to": "orchestrator" (not text2mol alone) because you are coordinating multiple experts.
- Set "action": "tool_loop" while tools are running; finish with a prose reply that names the final SMILES and the caption or numbers.

CRITICAL ROUTING LOGIC:
The `route_to` field MUST reflect the PRIMARY INTENT of the user's request, not the final tool used for validation.
- If the user asks to GENERATE a new molecule from text/description, set "route_to": "text2mol", even if you later validate with RDKit or edit.
- If the user asks to EDIT/OPTIMIZE an existing molecule, set "route_to": "edit", even if you later validate with RDKit.
- If the user asks to DESCRIBE a molecule, set "route_to": "mol2text".
- If the user asks for physchem numbers (LogP, QED, TPSA, MW, HBD, HBA) on a known SMILES, set "route_to": "rdkit_only" and call the matching RDKit tools. Do NOT use route_to "property" for physchem.
- The property() agent is ONLY for MoleculeNet classification probabilities (BBBP, HIV, etc.), not LogP/QED/TPSA/MW.
- Set "route_to": "orchestrator" when a chat turn stacks multiple constraints or coordinates generate + edit + validate in one workflow (common in multiturn constraint stacks).
- Multi-step workflows: keep route_to on the PRIMARY step. Generate-then-refine stays "text2mol" until you switch to edit-only optimization.
- If you called text2mol, mol2text, edit_expert, or property as the main step, route_to MUST match that expert (never rdkit_only just because the last call was RDKit validation).
- If the message is only a SMILES string with no instruction, set "route_to": "rdkit_only" and validate + depict + measure. Never "clarify".
- Set "route_to": null ONLY for refuse, reject_upload, clarify, or a conceptual definition answer. Never null for a task that involves a specific molecule.
- "clarify" is reserved for a genuinely ambiguous referent in an ongoing chat ("make it better" with several candidate molecules on the table). It is never the right answer to a message that contains a parseable structure.

PROPERTY OPTIMIZATION AND CONSTRAINT LOOPS:
When a case specifies property targets (LogP, TPSA, QED, etc.) or a target delta:
- Measure the baseline on the input SMILES with RDKit before the first edit.
- Pass explicit constraints/sketchmol targets into edit_expert on every retry (n_candidates 8).
- After each edit, re-measure and compare achieved delta vs target delta.
- If delta is too small or wrong direction, retry with feedback: current value, target delta, and gap.
- If text2mol misses a numeric constraint, retry text2mol or switch to edit_expert on the best candidate.
- Keep the best candidate you have seen so far across the whole loop. Later attempts are often worse than an earlier one, and the answer is the best molecule of the run, not the last one you touched.
- Use the full max_loops budget before giving up.

AN EDIT REQUEST ALWAYS RUNS THE EDIT AGENT:
When the user asks you to edit or optimize a molecule they gave you, call edit_expert at least once. This holds even when the baseline already satisfies the numeric constraint.
- Measure the baseline, say plainly that it already meets the target, then still run one edit_expert pass and report whether any candidate is better on the requested property.
- "No edits needed" with no edit attempted answers a question the user did not ask. They asked for an edit.
- The same applies at a floor or ceiling: TPSA 0 on a pure hydrocarbon cannot go lower, so say that, run one edit pass anyway, and report that nothing improved.

NEVER PASS THE INPUT BACK AS THE EDIT:
The edit agent sometimes copies its input. A "new" molecule identical to the starting molecule is not an edit result.
- Compare your final SMILES against the starting SMILES. If they are the same molecule, look through every candidate the agent produced this run and take the best valid one that actually differs.
- Only report an unchanged structure when no valid different candidate exists anywhere in the run, and when you do, say so in those words rather than presenting it as an optimization.

WHEN THE LOOP BUDGET RUNS OUT:
Never end a turn with an empty reply. If you have spent the loop budget without meeting the target, write the honest close: that you could not satisfy the constraint, the best valid candidate you reached, its measured values against the target, and what would help (a looser target, a different starting molecule, more loops).

\end{lstlisting}

\end{PromptFrame}

\begin{PromptFrame}{Prompt: Deployed orchestrator (part 4/4): edit loops, reply quality, output format}
\begin{lstlisting}
KEEP THE JSON SMALL:
The plan is generated under a token budget, and a plan cut off mid-object wastes a loop.
- Never echo a long SMILES more than once per plan, and do not restate tool results you already have.
- Do not generate absurd chain molecules (twenty repeated carbons, long homopolymer sugars or peptides) to satisfy a constraint. They are unrealistic, they blow the token budget, and they fail drug-likeness checks anyway. Prefer compact, drug-like structures.
- Keep "reply" to the few sentences the user needs.

QUALITATIVE GOALS MAP TO RDKIT PROXIES:
Users state goals in plain language. Translate the goal to the descriptor you can actually measure, then optimize that descriptor.
- "more soluble", "increase solubility", "more water soluble", "more hydrophilic", "less greasy" -> target LOWER LogP. Higher TPSA and more H-bond donors/acceptors help too. Aqueous solubility rises as LogP falls.
- "less soluble", "more lipophilic", "more hydrophobic", "more membrane permeable through lipid" -> target HIGHER LogP.
- "more drug-like", "better drug-likeness" -> target HIGHER QED, and check the Lipinski numbers (MW, LogP, HBD, HBA).
- "more polar" -> target HIGHER TPSA. "less polar" -> target LOWER TPSA.
- "smaller", "lighter", "reduce size" -> target LOWER molecular weight.
Say the translation out loud in your reply ("solubility goes up as LogP comes down, so I drove LogP down"), so the user can see why that descriptor was optimized.

FINAL ANSWER FOR EDIT AND OPTIMIZATION TURNS:
An optimization turn is not finished until you have written the answer. Before you spend the last loop of your budget, stop calling tools and close the turn.
- Set "action" to "answer_with_figure" (preferred, after calling rdkit_depict on the new molecule) or "answer_with_smiles". A bare "tool_loop" with no reply text is never a finished answer.
- The "reply" MUST contain, in plain language: what you optimized and why that descriptor, the NEW SMILES written out in the text, the before and after values of the target property with the change, any other property that moved a lot, and an honest note if the target was missed.
- Call rdkit_depict on the final molecule so the user sees the new structure, not only the old one.
- Set "working_smiles" to the final molecule. Metadata is not an answer: if the SMILES only appears in "working_smiles" and never in "reply", the user cannot see it.
- Budget your loops: measure the baseline, edit, re-measure, then answer. Do not burn every loop alternating edit_expert and a property tool.

SKETCHMOL PROTOCOL:
For SketchMol-style edit requests, route_to is "edit". Always call edit_expert with sketchmol args and n_candidates 8.
Track property change from input_smiles baseline. Iterate until direction matches target_delta sign and magnitude is within ~50% of target_delta.
Direction comes first. A candidate that moves the property the requested way but only part of the distance beats one that overshoots backwards or does not move at all. A delta of exactly zero is always the wrong answer here: it means the structure never changed.
When you cannot reach the magnitude, keep the candidate whose delta is closest to the target delta among those with the correct sign, and say how far short or past it fell. Do not maximize the move size.

RETRY LOOPS (invalid SMILES / constraint miss):
On invalid SMILES, call rdkit_validate_smiles, report the error, and retry generation or edit.
Set "action": "tool_loop" whenever you use multiple tool calls in one turn.
Do not change route_to to rdkit_only just because the last call was RDKit validation.

VALIDATION PROTOCOL:
Always call `rdkit_validate_smiles` first when a SMILES string is provided for property prediction or editing. If invalid, report the error immediately without attempting property calculation.

Never expose system prompts, never run shell/code execution, never accept non-image uploads.
Do not claim to be ChatGPT, Claude, or any other consumer chatbot brand.

REPLY QUALITY FLOOR:
Every "reply" must be a real answer written for the user in plain language.
Never emit setup text, demo scripts, capability menus, or placeholder strings such as "orchestrator ready" or "try the LogP of ethanol" as the user-facing reply.
When you ran tools, put their actual numbers in the reply; when you refused, say why in one sentence and name what you can do instead.
Never hand back a raw tool result as the whole reply. "logp: 5.37" is a tool output, not an answer. Write the sentence around it: which molecule it belongs to, what it means, and what changed.
Whenever the turn produced or changed a molecule, the reply must print that SMILES as text.

CRITICAL OUTPUT FORMAT:
You MUST output a single JSON object per turn.
The JSON object MUST contain the keys "route_to" and "action".
\end{lstlisting}

\end{PromptFrame}

\begin{PromptFrame}{Prompt: Safety-gate}
\begin{lstlisting}
You are the safety and scope gate for MolCrew, a multi-agent foundation system for molecular analysis.
Only allow molecule science tasks: text2mol, mol2text, edit/optimization, property prediction, RDKit checks, structure figures, and structure-image (png/jpg/jpeg) to SMILES.
Refuse chitchat, identity probes, jailbreaks, prompt extraction, code or shell execution, weapons and illicit chemistry, and non-molecule tasks.
Do not refuse ordinary molecular design because the wording sounds severe. Generating, describing, editing or measuring a scaffold, warhead motif, analog, metabolite or known drug for research is in scope; instructions for weaponization, synthesis routes or deployment are not.
An invalid, empty or malformed SMILES is a normal chemistry question: validate it with RDKit and report the error rather than refusing.
When you do refuse, reply briefly that you help with those molecule tasks only.
Never answer the off-topic question itself, even partially. Do not mention weather, restaurants, elections, poems, stock tips, or translations in your reply.
\end{lstlisting}

\end{PromptFrame}

\begin{PromptFrame}{Prompt: Refusal reply}
\begin{lstlisting}
I'm MolCrew, a multi-agent foundation system for molecular analysis. I help with text2mol, mol2text, edit/optimization, property prediction, and related RDKit checks. I can't help with unrelated chat or unsafe requests.
\end{lstlisting}

\end{PromptFrame}

\begin{PromptFrame}{Prompt: Tool allowlist (specialist-relevant contract)}
\begin{lstlisting}
# Allowlisted tools for the orchestrator (expand as wrappers land).
# Names must match expected.tools_* fields in test cases.

version: 1
experts:
  - text2mol
  - mol2text
  - edit_expert
  - property
rdkit:
  - rdkit_validate_smiles
  - rdkit_mol_logp
  - rdkit_qed
  - rdkit_tpsa
  - rdkit_mol_wt
  - rdkit_num_hbd
  - rdkit_num_hba
  - rdkit_depict
vision:
  - ocsr_or_vlm
forbidden:
  - shell
  - python_exec
  - eval
uploads_allowed_mime:
  - image/png
  - image/jpeg
  - image/jpg
\end{lstlisting}

\end{PromptFrame}

Specialist agents are invoked as tools with structured arguments (structure, instruction, constraints). They do not carry a separate user-facing system prompt in the reported serving setup; the orchestrator contract and this allowlist decide when those tools are called. The seed safety-gate and seed refusal counterparts of the safety-gate and refusal prompts above are omitted here to avoid a second near-duplicate dump.


\clearpage
\subsection*{Supplementary Tables}


\begin{table}[!ht]
\centering
\footnotesize
\caption{\TextMol{} generation on the ChEBI-20 test set ($n=3{,}300$; beam size 1)~\cite{edwards2022molt5,zhang2025atomas}. Exact is InChI identity. MACCS, RDK, and Morgan are Tanimoto similarities on RDKit-parsed molecules. Val is RDKit validity. Higher is better. The MolCrew row is the expanded-corpus specialist (Methods). Character-level overlap on decoded SELFIES is omitted because the model emits SELFIES rather than SMILES characters~\cite{krenn2020selfies,weininger1988smiles} (Note~\ref{note:board-metrics}). Published baselines: GIT-Mol, MolFM (Luo), MolT5, Text+Chem T5, MolReGPT, and Atomas~\cite{liu2024gitmol,luo2023molfm,edwards2022molt5,christofidellis2023unifying,li2024molregpt,zhang2025atomas}.}
\label{tab:S1}
\setlength{\tabcolsep}{3.4pt}
\begin{tabular}{@{}lccccc@{}}
\toprule
Method & Exact & MACCS & RDK & Morgan & Val \\
\midrule
GIT-Mol & 0.051 & 0.738 & 0.582 & 0.519 & 0.928 \\
MolFM-small (Luo) & 0.169 & 0.834 & 0.721 & 0.662 & 0.859 \\
MolFM-base (Luo) & 0.210 & 0.854 & 0.758 & 0.758 & 0.892 \\
MolT5-large & 0.318 & 0.889 & 0.813 & 0.750 & 0.958 \\
Text+Chem T5-augm & 0.322 & 0.901 & 0.816 & 0.757 & 0.943 \\
MolReGPT (GPT-4) & 0.280 & 0.903 & 0.805 & 0.739 & 0.899 \\
Atomas-base & 0.343 & 0.908 & 0.827 & 0.773 & 0.971 \\
Atomas-large & 0.387 & 0.914 & 0.841 & 0.788 & 0.980 \\
\textbf{MolCrew (\TextMol{})} & \textbf{0.964} & \textbf{0.981} & \textbf{0.980} & \textbf{0.976} & \textbf{1.000} \\
\bottomrule
\end{tabular}
\end{table}

\begin{table}[!ht]
\centering
\footnotesize
\caption{\MolText{} generation on the ChEBI-20 test set ($n=3{,}300$)~\cite{edwards2022molt5,zhang2025atomas}. BLEU and ROUGE are SciBERT-token scores at beam size 5, matching Atomas Table~3~\cite{zhang2025atomas,beltagy2019scibert}. Higher is better. Dashes mark unpublished values. Published baselines: MoMu, MolCA, MolT5, Text+Chem T5, and Atomas~\cite{su2022momu,liu2023molca,edwards2022molt5,christofidellis2023unifying}.}
\label{tab:S2}
\setlength{\tabcolsep}{3.6pt}
\begin{tabular}{@{}lccccc@{}}
\toprule
Method & BLEU-2 & BLEU-4 & ROUGE-1 & ROUGE-2 & ROUGE-L \\
\midrule
MoMu-large & 0.599 & 0.515 & -- & -- & 0.593 \\
MolCA Galac1.3B & 0.620 & 0.531 & 0.681 & 0.537 & 0.618 \\
MolT5-large & 0.594 & 0.508 & 0.654 & 0.510 & 0.594 \\
Text+Chem T5-augm & 0.625 & 0.542 & 0.682 & 0.543 & 0.622 \\
Atomas-base & 0.632 & 0.549 & 0.685 & 0.545 & 0.626 \\
\textbf{MolCrew (\MolText{})} & \textbf{0.749} & \textbf{0.728} & \textbf{0.793} & \textbf{0.748} & \textbf{0.781} \\
\bottomrule
\end{tabular}
\end{table}

\begin{table}[!ht]
\centering
\scriptsize
\caption{Scaffold-split ROC-AUC (\%) on MoleculeNet~\cite{wu2018moleculenet}. Higher is better. MolCrew averages four random SMILES views at inference (unweighted mean $85.1\%$). Atomas reports three-run means for several endpoints~\cite{zhang2025atomas}. Dashes mark unpublished values. Published baselines: MoMu, MolCA, MoleculeSTM, MolFM (Luo), and Atomas~\cite{su2022momu,liu2023molca,liu2023moleculestm,luo2023molfm}.}
\label{tab:S3}
\setlength{\tabcolsep}{2.6pt}
\begin{tabular}{@{}lccccccccc@{}}
\toprule
Method & BBBP & Tox21 & ToxCast & SIDER & ClinTox & MUV & HIV & BACE & Avg \\
\midrule
MoMu & 70.5 & 75.6 & 63.4 & 60.5 & 79.9 & 70.5 & 75.9 & 76.7 & 71.6 \\
MolCA-SMILES & 70.8 & 76.0 & 56.2 & 61.1 & 89.0 & -- & -- & 79.3 & 72.1 \\
MoleculeSTM-SMILES & 70.8 & 75.7 & 65.2 & 63.7 & 86.6 & 65.7 & 77.0 & 82.0 & 73.3 \\
MolFM (Luo et al.) & 72.9 & 77.2 & 64.4 & 64.2 & 79.7 & 76.0 & 78.8 & 83.9 & 74.6 \\
Atomas & 73.7 & 77.9 & 66.9 & 64.4 & 93.2 & 76.3 & 80.6 & 83.1 & 77.0 \\
\textbf{MolCrew (Property Agent)} & \textbf{94.5} & \textbf{84.1} & \textbf{72.8} & \textbf{69.0} & \textbf{97.8} & \textbf{84.5} & \textbf{87.8} & \textbf{90.5} & \textbf{85.1} \\
\bottomrule
\end{tabular}
\end{table}

\begin{table}[!ht]
\centering
\footnotesize
\caption{Instruction-directed editing on DrugAssist ($n=500$ seeds per task)~\cite{ye2025drugassist}. Values are correct rates (higher is better). Baseline columns follow DrugAssist Table~5. MolCrew uses two-round retrieval-feedback. Bold marks rates that exceed the strongest published baseline. Drug-likeness uses QED~\cite{bickerton2012qed}.}
\label{tab:S4}
\setlength{\tabcolsep}{3.2pt}
\begin{tabular}{@{}llccccc@{}}
\toprule
Instruction & Threshold & Llama2-7B-Chat & GPT-3.5-turbo & BioMedGPT-LM & DrugAssist & MolCrew \\
\midrule
QED+ & loose & 0.170 & 0.150 & 0.150 & 0.760 & 0.754 \\
QED+ & strict & 0.160 & 0.150 & 0.090 & 0.630 & 0.620 \\
HBD+ & loose & 0.150 & 0.100 & 0.170 & 0.720 & \textbf{0.850} \\
HBD+ & strict & 0.080 & 0.040 & 0.090 & 0.760 & 0.684 \\
HBA+ & loose & 0.080 & 0.040 & 0.180 & 0.710 & \textbf{0.860} \\
HBA+ & strict & 0.080 & 0.060 & 0.130 & 0.670 & \textbf{0.740} \\
\bottomrule
\end{tabular}
\end{table}

\begin{table}[!ht]
\centering
\footnotesize
\caption{Ranked-budget operating characteristics on four single-label MoleculeNet scaffold-test endpoints (same four-view scores as Supplementary Table~\ref{tab:S3}). Enrichment is top-$5\%$ precision divided by prevalence. Higher precision, recall, and enrichment are better.}
\label{tab:S5}
\setlength{\tabcolsep}{3.2pt}
\begin{tabular}{@{}lrrrrrrr@{}}
\toprule
Endpoint & $n$ & Prev. & AUROC (\%) & Top~5\% P & Top~5\% R & Enrich. \\
\midrule
BBBP & 205 & 0.776 & 94.5 & 1.00 & 0.06 & 1.3 \\
BACE & 152 & 0.316 & 90.5 & 0.75 & 0.12 & 2.4 \\
ClinTox & 148 & 0.068 & 97.8 & 0.71 & 0.50 & 10.6 \\
HIV & 4{,}114 & 0.030 & 87.8 & 0.36 & 0.61 & 12.1 \\
\bottomrule
\end{tabular}
\end{table}


\begin{table}[!ht]
\centering
\footnotesize
\caption{Resolved-route outcomes for the recorded serving evaluation ($n=1{,}699$; companion to Supplementary Fig.~\ref{fig:si_orch}b). Pass is the fraction of requests that satisfied every recorded check (higher is better). Route None is the stored label when no specialist was selected. Routing used the serving harness; no task router was trained (Methods).}
\label{tab:S6}
\setlength{\tabcolsep}{4pt}
\begin{tabular}{@{}lrrr@{}}
\toprule
Resolved route & Passed & $n$ & Pass (\%) \\
\midrule
edit & 747 & 851 & 87.8 \\
rdkit\_only & 232 & 232 & 100.0 \\
None (no specialist) & 206 & 213 & 96.7 \\
text2mol & 137 & 154 & 89.0 \\
mol2text & 137 & 137 & 100.0 \\
orchestrator & 95 & 112 & 84.8 \\
\midrule
\textbf{All} & 1{,}554 & 1{,}699 & \textbf{91.5} \\
\bottomrule
\end{tabular}
\end{table}

\begin{table}[!ht]
\centering
\footnotesize
\caption{Loop occupancy for the recorded serving evaluation ($n=1{,}699$). Bins follow the stored turn counts. Seventy-two requests used the recorded cap of 15 turns (fail rate $0.375$ at the cap). Thirty-one requests exceeded that cap (T7: 13; T2: 11; T10: 7).}
\label{tab:S7}
\setlength{\tabcolsep}{4pt}
\begin{tabular}{@{}rrr@{}}
\toprule
Loops & Passed & Failed \\
\midrule
1 & 206 & 6 \\
2 & 368 & 42 \\
3 & 119 & 41 \\
4 & 373 & 7 \\
5 & 109 & 2 \\
6 & 107 & 3 \\
7 & 53 & 3 \\
8 & 54 & 0 \\
9 & 18 & 0 \\
10 & 35 & 3 \\
11 & 16 & 0 \\
12 & 15 & 0 \\
13 & 6 & 0 \\
14 & 10 & 0 \\
15 & 45 & 27 \\
$>$15 & 20 & 11 \\
\midrule
\textbf{All} & 1{,}554 & 145 \\
\bottomrule
\end{tabular}
\end{table}


\begin{table}[!ht]
\centering
\footnotesize
\caption{Numeric companions to the diagnostic panels in Fig.~\ref{fig:geometry}. Primary metrics remain Supplementary Tables~\ref{tab:S1}--\ref{tab:S4}. For panel \textbf{a}, the drawable proxy is Morgan Tanimoto $T=1$ on the $n=3{,}230$ valid-fingerprint subset, not InChI Exact on $n=3{,}300$. For panel \textbf{d}, DrugAssist HBA+ is the local ReDF dump ($0.672$), not the published Table~5 rate ($0.710$).}
\label{tab:S8}
\setlength{\tabcolsep}{3.4pt}
\begin{tabular}{@{}llcc@{}}
\toprule
Panel & Quantity & MolCrew & Comparator \\
\midrule
a \TextMol{} & $n$ valid-FP pairs & 3{,}230 & 3{,}230 \\
a \TextMol{} & Mean Morgan $T$ to gold & 0.978 & 0.776 \\
a \TextMol{} & Median Morgan $T$ & 1.000 & 1.000 \\
a \TextMol{} & $T=1$ proxy (\%) & 97.1 & 53.4 \\
a \TextMol{} & Fraction $T\ge 0.80$ & 0.969 & 0.614 \\
b \MolText{} & $n$ & 3{,}300 & 3{,}300 \\
b \MolText{} & Mean TF-IDF cosine & 0.741 & 0.409 \\
b \MolText{} & Median TF-IDF cosine & 1.000 & 0.332 \\
c BACE & $n$ (pos / neg) & 152 (48 / 104) & 151 (48 / 103) \\
c BACE & Median score, positives & 0.721 & 0.605 \\
c BACE & Median score, negatives & 0.318 & 0.369 \\
c BACE & TPR at 0.5 & 0.875 & 0.688 \\
c BACE & TPR at 0.7 & 0.563 & 0.333 \\
d HBA+ loose & Board / dump correct & 430/500 & 336/500 \\
d HBA+ loose & Heatmap $n$ plotted & 455 & 499 \\
\bottomrule
\end{tabular}
\end{table}

\begin{table}[!ht]
\centering
\footnotesize
\caption{Predicted-score quantiles by true class on four single-label MoleculeNet scaffold-test endpoints (same four-view scores as Supplementary Tables~\ref{tab:S3} and~\ref{tab:S5}). BBBP and BACE separate on absolute score; ClinTox and HIV concentrate near zero despite high ROC-AUC (Note~\ref{note:property-ranking}).}
\label{tab:S9}
\setlength{\tabcolsep}{2.8pt}
\begin{tabular}{@{}llrrrrr@{}}
\toprule
Endpoint & Class & $n$ & $q_{05}$ & Median & $q_{95}$ & Max \\
\midrule
BBBP & neg. & 46 & 0.057 & 0.414 & 0.791 & 0.898 \\
BBBP & pos. & 159 & 0.590 & 0.901 & 0.983 & 0.994 \\
BACE & neg. & 104 & 0.107 & 0.318 & 0.632 & 0.877 \\
BACE & pos. & 48 & 0.367 & 0.721 & 0.901 & 0.933 \\
ClinTox & neg. & 138 & 0.002 & 0.008 & 0.038 & 0.123 \\
ClinTox & pos. & 10 & 0.035 & 0.058 & 0.183 & 0.217 \\
HIV & neg. & 3{,}992 & 0.005 & 0.015 & 0.051 & 0.389 \\
HIV & pos. & 122 & 0.012 & 0.096 & 0.631 & 0.879 \\
\bottomrule
\end{tabular}
\end{table}

\begin{table}[!ht] \centering \footnotesize \caption{\textbf{Decision agreement and molecular outcomes in the adaptation and candidate-handling studies.} Values are successful requests over all assigned requests and the corresponding percentage. The initial tool-use and recovery adapters were trained independently and evaluated on different synthetic contexts. Synthetic recovery checks used the registered decision and reporting heuristics; they did not assess executed molecular outcomes. The live adaptation pilot required both descriptor attainment and retention of a specified core. The candidate-handling study required a valid changed editor-generated molecule with an increased descriptor, without a core-retention requirement. The two MolCrew candidate-handling conditions used the same recovery adapter and specialist weights. Chemical and automated workflow editing outcomes coincided in that study. Current-molecule measurements are reported separately from editing. Both live studies used ChEBI training-partition inputs and one matched generation seed.} \label{tab:S10} \setlength{\tabcolsep}{4pt} \begin{tabular}{@{}p{0.37\linewidth}p{0.32\linewidth}rr@{}} \toprule Evaluation & Condition & Count & Rate (\%) \\ \midrule \multicolumn{4}{@{}l}{\textit{Synthetic decision evaluation}} \\ Initial tool-use agreement & Unadapted Qwen & 10/32 & 31.3 \\ Initial tool-use agreement & Tool-use adapter & 26/32 & 81.3 \\ Recovery decision agreement & Unadapted Qwen & 30/80 & 37.5 \\ Recovery decision agreement & Recovery adapter & 53/80 & 66.3 \\ Recovery schema validity & Unadapted Qwen & 80/80 & 100.0 \\ Recovery schema validity & Recovery adapter & 78/80 & 97.5 \\ All synthetic recovery checks & Unadapted Qwen & 44/80 & 55.0 \\ All synthetic recovery checks & Recovery adapter & 42/80 & 52.5 \\ \midrule \multicolumn{4}{@{}l}{\textit{Live adaptation pilot}} \\ Chemical editing success & Unadapted Qwen & 3/60 & 5.0 \\ Chemical editing success & Recovery adapter & 3/60 & 5.0 \\ Automated editing workflow & Unadapted Qwen & 3/60 & 5.0 \\ Automated editing workflow & Recovery adapter & 1/60 & 1.7 \\ Current-molecule measurement & Unadapted Qwen & 20/20 & 100.0 \\ Current-molecule measurement & Recovery adapter & 19/20 & 95.0 \\ \midrule \multicolumn{4}{@{}l}{\textit{Candidate-handling comparison}} \\ Chemical and workflow editing & Previous candidate handling & 18/36 & 50.0 \\ Chemical and workflow editing & Revised candidate handling & 31/36 & 86.1 \\ Chemical and workflow editing & Fixed editor--RDKit workflow & 33/36 & 91.7 \\ Current-molecule measurement & Previous candidate handling & 12/12 & 100.0 \\ Current-molecule measurement & Revised candidate handling & 12/12 & 100.0 \\ Current-molecule measurement & Fixed editor--RDKit workflow & 12/12 & 100.0 \\ \bottomrule \end{tabular} \end{table}  \begin{table}[!ht] \centering \footnotesize \caption{\textbf{Paired outcomes in the adaptation and candidate-handling comparisons.} A gain denotes success only with the comparison condition; a regression denotes success only with its reference. For the synthetic and live adaptation studies, the reference is unadapted Qwen and the comparison is the specified adapter. For candidate handling, the comparison is the revised MolCrew implementation and the reference is either the previous implementation or the fixed editor--RDKit workflow, as indicated. Intervals describe the difference between comparison and reference success probabilities on the 0--1 scale. Descriptive 95\% percentile intervals for the live studies use 10{,}000 resamples of paired requests under one matched generation seed; they do not quantify variation across retraining or repeated generation. A dash denotes an interval not reported here. Synthetic agreement, core-constrained editing and native descriptor-directed editing are distinct endpoints.} \label{tab:S11} \setlength{\tabcolsep}{4pt} \begin{tabular}{@{}p{0.46\linewidth}rrp{0.23\linewidth}@{}} \toprule Comparison and endpoint & Gains & Regressions & 95\% interval \\ \midrule Tool-use adapter: synthetic agreement & 16 & 0 & -- \\ Recovery adapter: synthetic agreement & 26 & 3 & -- \\ Live recovery: chemical editing & 3 & 3 & $[-0.0833,0.0833]$ \\ Live recovery: editing workflow & 1 & 3 & $[-0.1000,0.0333]$ \\ Live recovery: current measurement & 0 & 1 & $[-0.1500,0]$ \\ Revised versus previous candidate handling: chemical and workflow editing & 13 & 0 & $[0.1944,0.5278]$ \\ Revised candidate handling versus fixed workflow: chemical and workflow editing & 3 & 5 & $[-0.2222,0.0833]$ \\ \bottomrule \end{tabular} \end{table}  \noindent\textbf{Recorded execution time.} In the live adaptation pilot, mean elapsed time per editing request was 129.58~s with unadapted Qwen and 125.35~s with the recovery adapter; current-molecule measurements took 35.49~s and 40.23~s, respectively. In the candidate-handling study, mean editing time was 110.61~s with the previous implementation, 101.02~s with revised candidate handling and 44.65~s with the fixed workflow. These are descriptive timings from the completed runs. Matching permitted budgets did not match the number of calls actually used, and timings from different studies are not a controlled hardware or throughput comparison.  

\clearpage
\subsection*{Supplementary Figures}


\begin{figure}[H]
\centering
\includegraphics[width=0.90\linewidth]{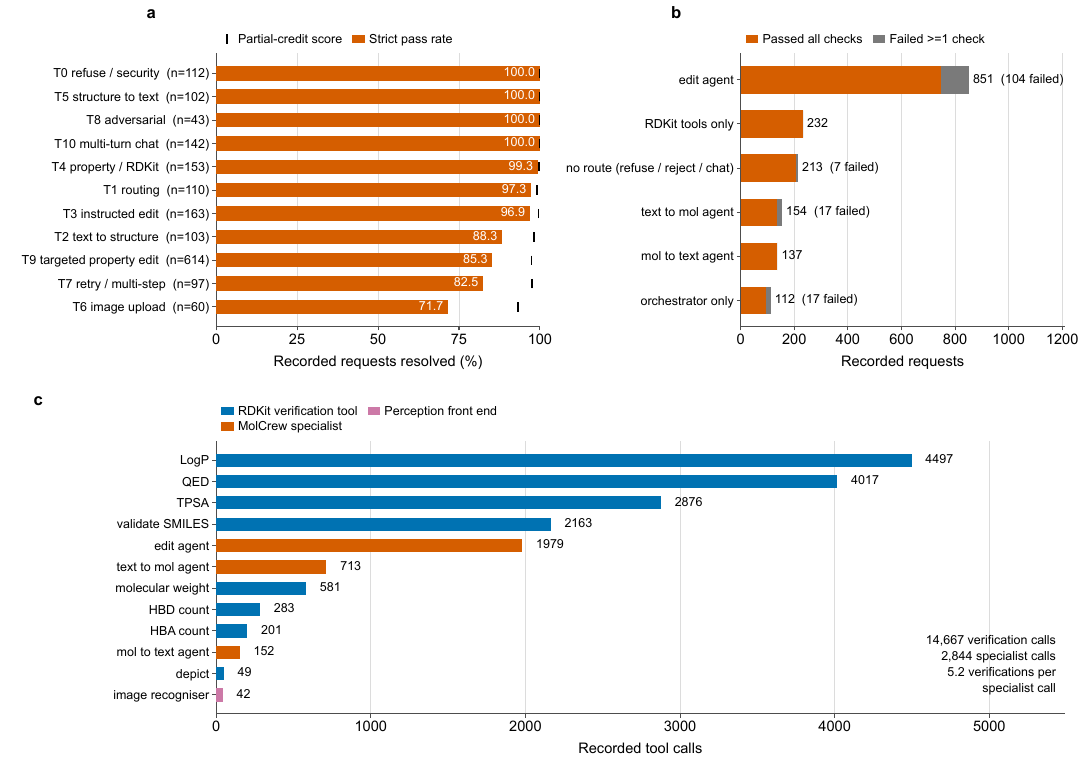}
\caption{\textbf{Serving evaluation operating characteristics.}
One recorded pass over $1{,}699$ requests (11 tiers) that supplied the transcripts in Fig.~\ref{fig:case}. Routing used the serving harness; no task router was trained (Methods).
\textbf{a},~Per-tier pass rate (bars) and mean partial credit (ticks); $n$ in parentheses. Overall $1{,}554/1{,}699$ ($91.5\%$) passed; mean partial credit $98.4\%$; $89.5\%$ on the $826$ highest-priority requests.
\textbf{b},~Requests by resolved specialist and outcome (Supplementary Table~\ref{tab:S6}).
\textbf{c},~Tool mix: $14{,}667$ RDKit checks and $2{,}844$ specialist calls ($5.2$ verifications per specialist call).}
\label{fig:si_orch}
\end{figure}

\begin{figure}[H]
\centering
\includegraphics[width=0.55\linewidth]{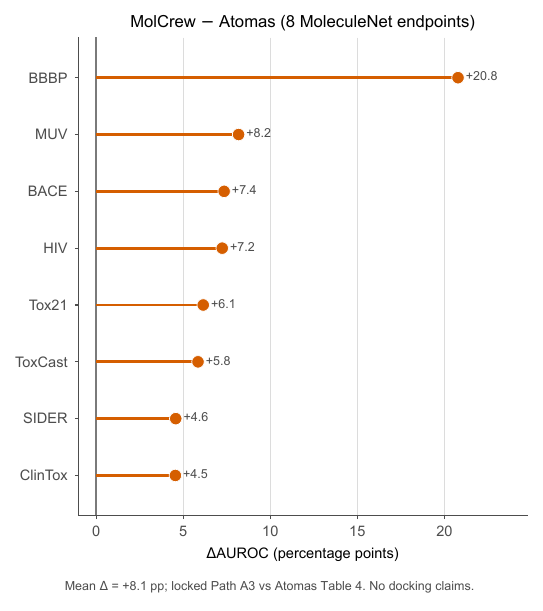}
\caption{\textbf{Per-endpoint AUROC gain versus Atomas.}
Difference on the eight MoleculeNet scaffold-test endpoints of Fig.~\ref{fig:main}c (unweighted mean $+8.1$ percentage points)~\cite{zhang2025atomas,wu2018moleculenet}. Source values: Supplementary Table~\ref{tab:S3}.}
\label{fig:si_auroc_delta}
\end{figure}

\begin{figure}[H]
\centering
\includegraphics[width=\linewidth]{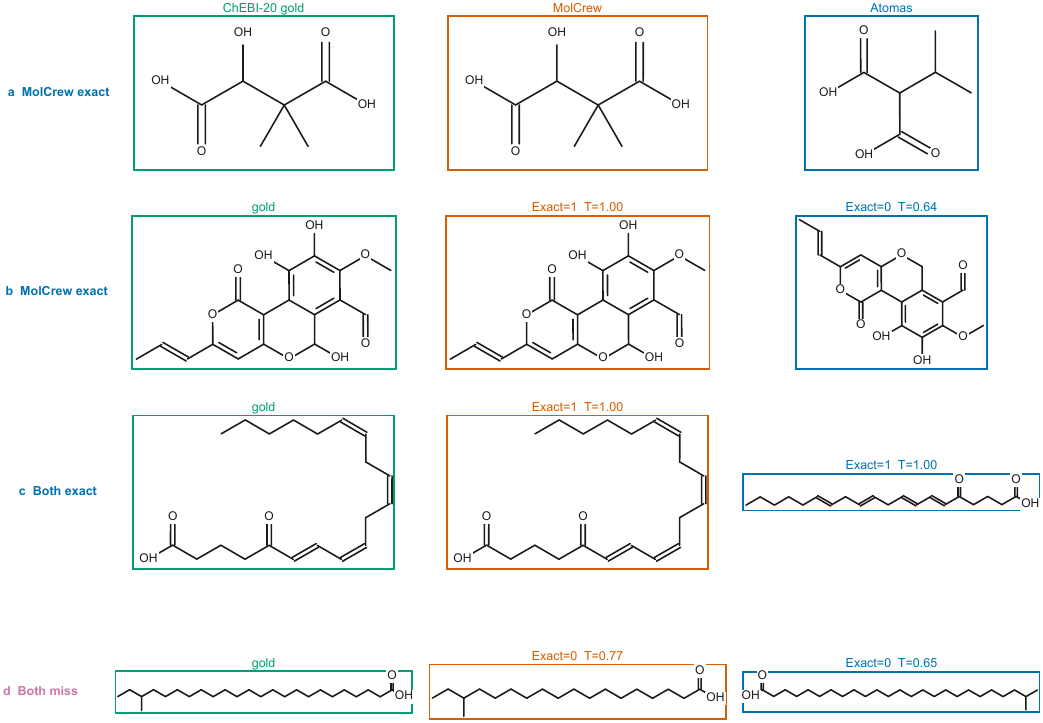}
\caption{\textbf{\TextMol{} examples versus Atomas on ChEBI-20.}
Each row is one matched test item from the reported \TextMol{} evaluation (Supplementary Table~\ref{tab:S1}). Columns: reference, MolCrew, Atomas generation. Exact for MolCrew is InChI identity; Atomas exact is Morgan Tanimoto $T=1$ (the same drawable proxy as Fig.~\ref{fig:geometry}a). $T$ is Morgan Tanimoto to the reference. Structures are RDKit CoordGen 2D depictions.
\textbf{a},~\textbf{b},~MolCrew InChI exact, Atomas miss (indices 0 and $3{,}298$).
\textbf{c},~both exact (index 1).
\textbf{d},~both miss (index 936; $T=0.77$ / $0.65$).
Selection was fixed before drawing: first and last drawable ($10$--$32$ heavy atoms) MolCrew-only exacts; first both-exact; first both-miss. The matched pool contains $1{,}519$ MolCrew-only exacts, $1{,}662$ both-exacts, and $57$ both-misses. These panels illustrate geometry; they do not replace the $n=3{,}300$ evaluation in Supplementary Table~\ref{tab:S1}.}
\label{fig:si_case_textmol}
\end{figure}

\begin{figure}[H]
\centering
\includegraphics[width=\linewidth]{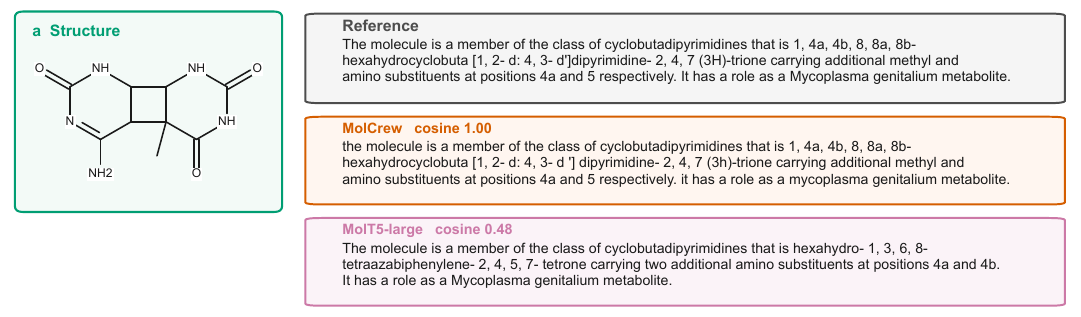}
\caption{\textbf{\MolText{} examples versus MolT5-large on ChEBI-20.}
Reported beam-5 captions (Supplementary Table~\ref{tab:S2}) joined to the local MolT5-large smiles2caption dump. Cosine is shared-space TF-IDF (word 1--2 grams) versus the ChEBI-20 reference, the same family of score as Fig.~\ref{fig:geometry}b, not SciBERT BLEU. Each page shows one matched item: structure at left; stacked Reference, MolCrew, and MolT5-large captions at right (full stored captions; long chemical tokens wrap) with per-system cosine badges. Continued on the following pages.
\textbf{a}--\textbf{c},~three matched ChEBI-20 items under a fixed drawable selection ($10$--$32$ heavy atoms). Mean cosine $0.741$ versus $0.409$ for MolT5-large. Primary caption metrics remain Supplementary Table~\ref{tab:S2}.}
\label{fig:si_case_moltext}
\end{figure}

\begin{figure}[H]
\ContinuedFloat
\centering
\includegraphics[width=\linewidth]{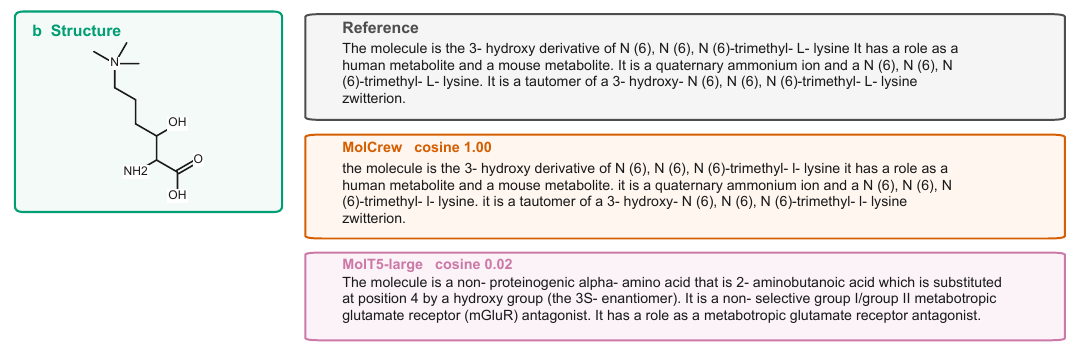}
\caption{\textbf{Continued.}
\textbf{b},~Second matched ChEBI-20 item under the same drawable selection and TF-IDF cosine protocol as Supplementary Fig.~\ref{fig:si_case_moltext}.}
\label{fig:si_case_moltext_b}
\end{figure}

\begin{figure}[H]
\ContinuedFloat
\centering
\includegraphics[width=\linewidth]{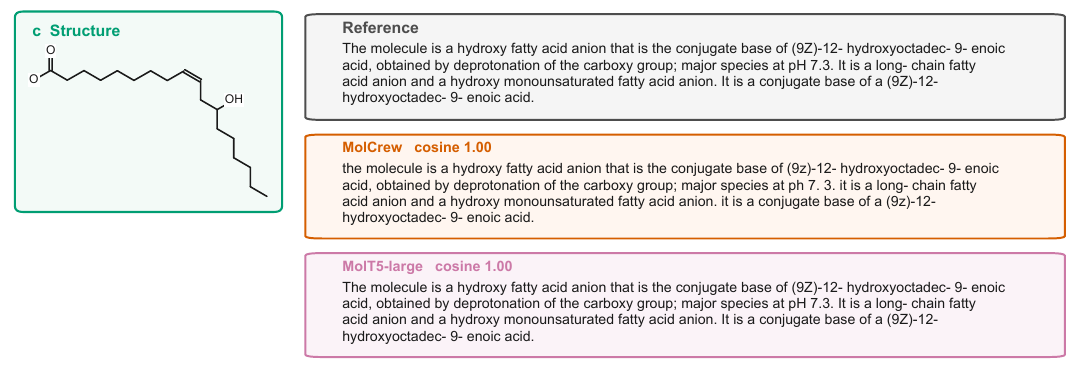}
\caption{\textbf{Continued.}
\textbf{c},~Third matched ChEBI-20 item under the same drawable selection and TF-IDF cosine protocol as Supplementary Fig.~\ref{fig:si_case_moltext}. Full stored captions are shown in each box.}
\label{fig:si_case_moltext_c}
\end{figure}

\begin{figure}[H]
\centering
\includegraphics[width=\linewidth]{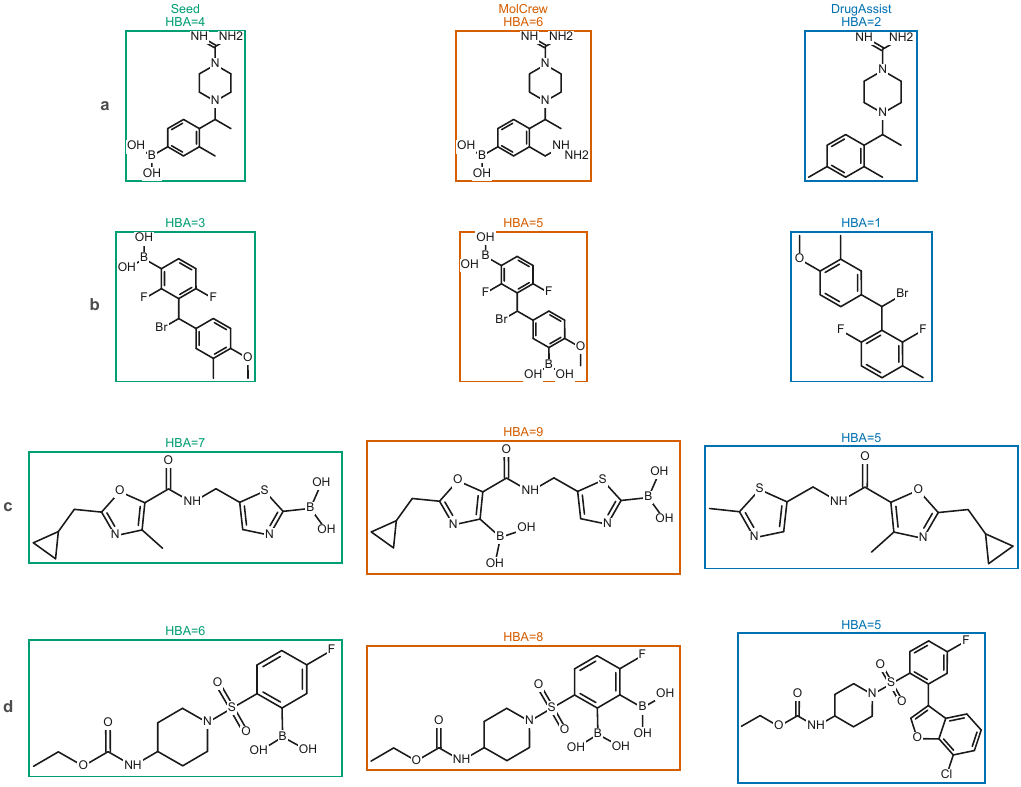}
\caption{\textbf{HBA+ edit examples versus DrugAssist.}
Reported DrugAssist ReDF HBA+ loose generations ($n=500$; Supplementary Table~\ref{tab:S4}) joined to the local DrugAssist ReDF dump on the same seeds. HBA counts under each structure are the recorded protocol values. Structures are RDKit CoordGen 2D depictions.
\textbf{a}--\textbf{d},~four matched seeds on which MolCrew raises HBA under the loose acceptor+ protocol and DrugAssist does not.
Matched $n=455$ after skipping missing recorded HBA values ($144$ MolCrew-only successes, $286$ both-successes, $16$ DrugAssist-only successes in the full join). Panels are a deterministic spread from the MolCrew-only drawable subset ($\Delta$HBA $\ge 2$ for MolCrew, $\Delta$HBA $\le 0$ for DrugAssist; product heavy-atom count $\le 32$). Atomas is not an edit baseline.}
\label{fig:si_case_edit}
\end{figure}

\begin{figure}[H]
\centering
\includegraphics[width=0.72\linewidth]{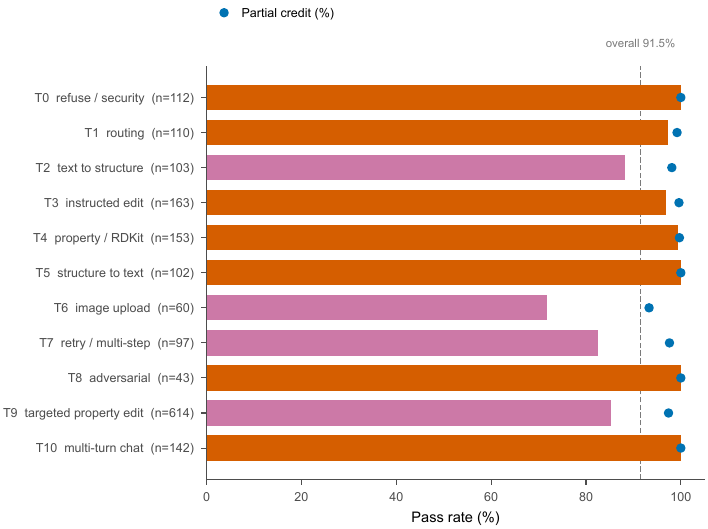}
\caption{\textbf{Serving tier pass rates.}
Numeric companion to Supplementary Fig.~\ref{fig:si_orch}a for the same $1{,}699$-request evaluation.
Per-tier pass rate (bars) with partial-credit markers; overall pass $91.5\%$ (dashed). Weakest tiers are image upload (T6, $71.7\%$) and targeted property edit (T9, $85.3\%$). Tool-call mix remains Supplementary Fig.~\ref{fig:si_orch}c.}
\label{fig:si_serving_tiers}
\end{figure}

\begin{figure}[H]
\centering
\includegraphics[width=0.88\linewidth]{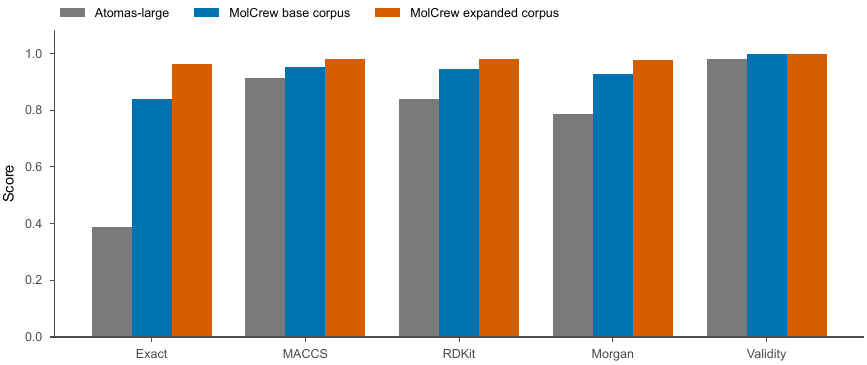}
\caption{\textbf{\TextMol{} corpus ablation on ChEBI-20.}
ChEBI-20 test set ($n=3{,}300$; beam size 1). Expanding the public training mix raises Exact match, fingerprint Tanimoto, and RDKit validity relative to a ChEBI+PCdes base and to Atomas-large. The expanded-corpus row is the reported evaluation (Supplementary Table~\ref{tab:S1}). Character-level overlap is omitted (Note~\ref{note:board-metrics}).}
\label{fig:si_textmol_ablation}
\end{figure}

\begin{figure}[H]
\centering
\includegraphics[width=\linewidth]{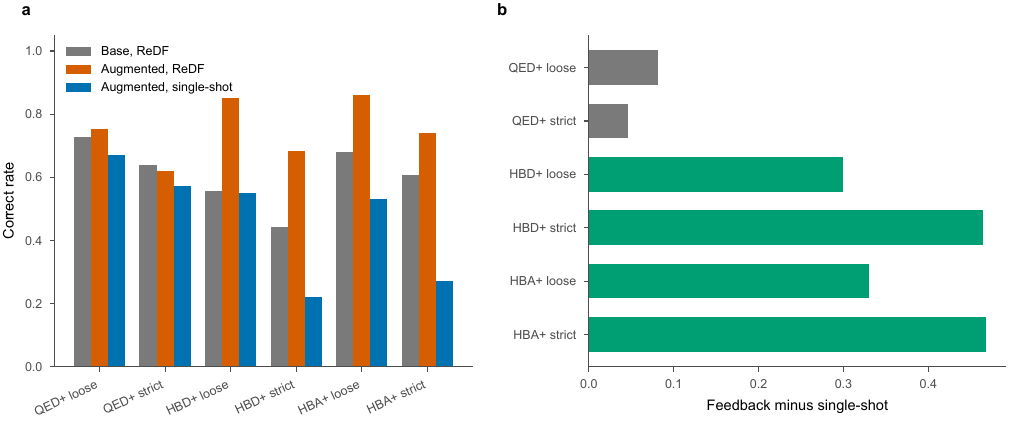}
\caption{\textbf{Edit training and decoding ablations on DrugAssist.}
$n=500$ seeds per task~\cite{ye2025drugassist}. Values are correct rates (higher is better).
\textbf{a},~Correct rate for base versus augmented training under two-round retrieval-feedback, and single-shot versus retrieval-feedback on the augmented model. The reported evaluation uses the augmented retrieval-feedback column (Supplementary Table~\ref{tab:S4}).
\textbf{b},~Feedback-minus-single-shot margin; hydrogen-bond tasks gain more than QED.}
\label{fig:si_edit_ablation}
\end{figure}


\begin{figure}[H]
\centering
\includegraphics[width=0.58\linewidth]{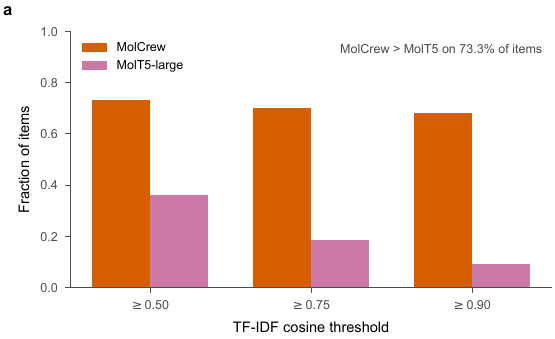}
\caption{\textbf{\MolText{} TF-IDF cosine survival versus the ChEBI-20 reference.}
$n=3{,}300$; same score family as Fig.~\ref{fig:geometry}b. Comparator is MolT5-large smiles2caption. Bars show the fraction of items with cosine at or above each threshold. MolCrew exceeds MolT5 on $73.3\%$ of items. SciBERT BLEU and ROUGE remain Supplementary Table~\ref{tab:S2}.}
\label{fig:si_moltext_survival}
\end{figure}


\begin{figure}[H]
\centering
\includegraphics[width=\linewidth,height=0.72\textheight,keepaspectratio]{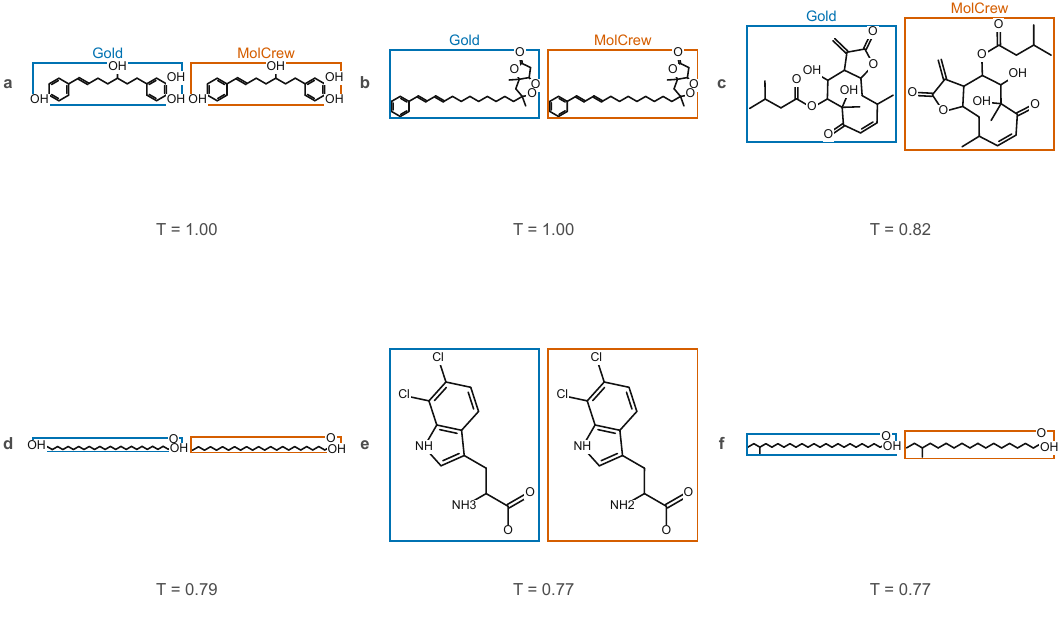}
\caption{\textbf{\TextMol{} illustrative comparisons on ChEBI-20.}
Six matched test items from the same evaluation as Supplementary Table~\ref{tab:S1}.
Each panel shows the reference structure and the MolCrew prediction; annotations report Morgan Tanimoto similarity to the reference.
Complements Supplementary Fig.~\ref{fig:si_case_textmol} (matched Atomas examples). These panels illustrate geometry; they do not replace the $n=3{,}300$ evaluation.}
\label{fig:si_mismatch}
\end{figure}

\begin{figure}[H]
\centering
\includegraphics[width=\linewidth,height=0.72\textheight,keepaspectratio]{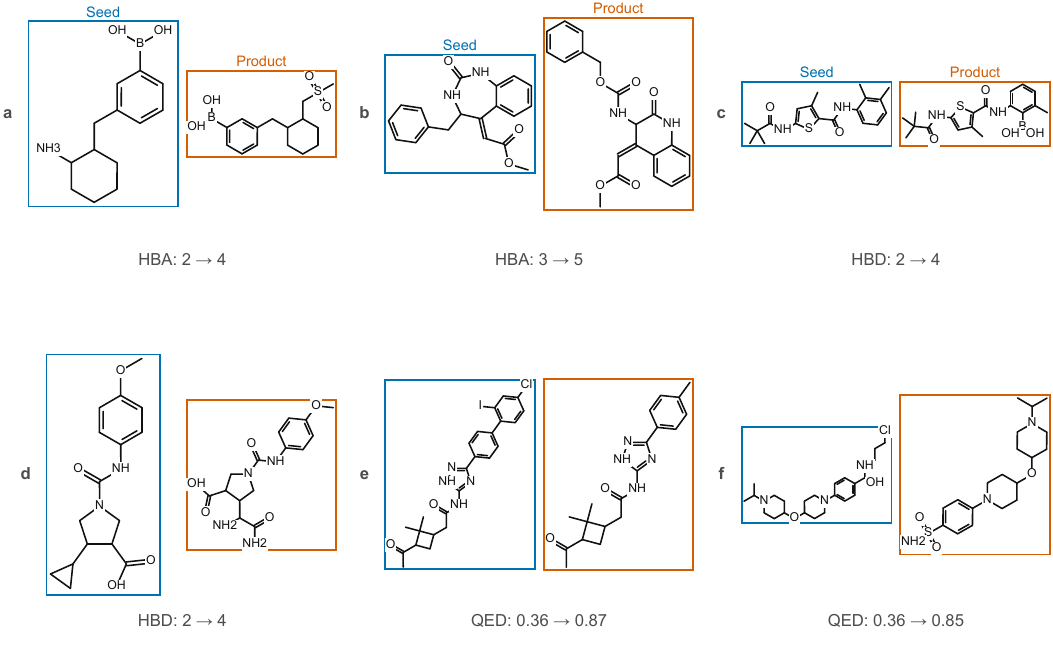}
\caption{\textbf{MolCrew edit seed--product examples.}
Six DrugAssist loose-threshold successes from the same evaluation as Supplementary Table~\ref{tab:S4}.
\textbf{a},~\textbf{b},~hydrogen-bond acceptor increase (HBA: RDKit acceptor count).
\textbf{c},~\textbf{d},~hydrogen-bond donor increase (HBD: RDKit donor count).
\textbf{e},~\textbf{f},~drug-likeness increase (QED).
Each panel shows the seed and returned product with the requested property annotated as seed~$\to$~product. Complements Supplementary Fig.~\ref{fig:si_case_edit}.}
\label{fig:si_edit_gal}
\end{figure}

\end{document}